\documentclass[aps,prl,amsmath,amssymb,reprint]{revtex4-1}
\usepackage[utf8]{inputenc}
\usepackage{mfirstuc}
\usepackage{graphicx}
\usepackage{subfigure}
\usepackage{float}
\usepackage{gensymb}
\usepackage{amsmath}
\usepackage{color}
\usepackage{multirow}
\usepackage{comment}
\usepackage[colorlinks=true, linkcolor=blue, citecolor=blue, urlcolor=blue]{hyperref}

\newcommand{\SK}[1]{\textcolor{black}{{#1}}}

\begin{document}
\title{Growth of Dynamic and Static Correlations in the Aging Dynamics of a Glass-Forming Liquid}

\author{Santu Nath}
\email{snath@tifrh.res.in}
\affiliation{Tata Institute of Fundamental Research, 36/P, Gopanpally Village, Serilingampally Mandal, Ranga Reddy District, Hyderabad 500046, Telangana, India}

\author{Smarajit Karmakar}
\email{smarajit@tifrh.res.in}
\affiliation{Tata Institute of Fundamental Research, 36/P, Gopanpally Village, Serilingampally Mandal, Ranga Reddy District, Hyderabad 500046, Telangana, India}
\date{\today}

\begin{abstract}
\SK{Using extensive molecular dynamics simulations, we have performed finite-size scaling (FSS) in the aging regime of a model glass-forming liquid to investigate how the length scales associated with amorphous order (static length) and dynamic heterogeneity (dynamic length) evolve with waiting time. The $\alpha$-relaxation time in the aging regime reveals non-monotonic finite-size effects with a peak at an intermediate system size, which, as far as we know, are not found in the equilibrium systems, and the peak position shifts to larger system sizes with decreasing temperature and increasing waiting time, indicating a growth of a characteristic length scale with waiting time. The extracted correlation volume associated with amorphous order increases logarithmically with the waiting time. Detailed analysis of the dependence of the length scale on waiting time allowed us to estimate the static length scale in the deep supercooled liquid regime. The dynamic length scale, obtained from FSS and block analysis of the four-point dynamic susceptibility, follows a power-law growth with waiting time. The values of the length scales obtained agree well with those obtained from different spatial correlation functions.}
\end{abstract}

\maketitle
\noindent{\bf \large Introduction: }\SK{Quenching a supercooled liquid well below the glass transition temperature from a high-temperature equilibrated state makes it out-of-equilibrium over a long time window. Unlike the situation in thermal equilibrium, measurements in the out-of-equilibrium states strongly depend on how much time has evolved; this phenomenon is known as \textit{Aging}~\cite{struik_1976,barrat_prl_1997,parisi_1997,barrat_phy-A_1999,parisi_1999,barrat_2000,berthier_book_2009,berthier_rmp_2011,saroj_prl_2012}. In an aging state, the physical properties of a glass are influenced by its waiting time, $t_w$, which allows us to tune its mechanical and optical properties by suitably tuning the preparation history, as demonstrated in a remarkable experiment on organic glasses prepared by two different ways~\cite{stephen_science_2007}. The waiting time dependence of the out-of-equilibrium dynamics has been extensively studied experimentally~\cite{lundgren_prl_1983,mattsson_1992,mattsson_prl_1995} and numerically~\cite{andersson_prb_1992,ritort_prb_1994,rieger_phys-A_1994,rieger_epl_1994,rieger_prb_1996} in various spin-glass models. Aging in structural glasses~\cite{barrat_prl_1997} was first time numerically investigated in the well-known Kob-Andersen model~\cite{kob_PRE_1995}. Traditionally aging in thermal glasses has been studied via temperature quenches; however, recent studies have shown that aging in active glasses~\cite{kallol_pnas_2023,subhodeep_2025} can be induced by both temperature and activity quenches~\cite{janzen_prr_2022, rituparno_prl_2020}}.

\SK{Unlike the equilibrium dynamics, time correlation functions in aging depend on both the time difference $t$ and the waiting time $t_w$. It exhibits a scaling behavior with the ratio $t/t_w^{\mu}$ ($\mu \le 1$) in the long time limit, and the corresponding relaxation time grows as $t_w^\mu$. Similar scaling analysis of the spatial correlation function indicates slow domain growth with increasing waiting time in spin glasses~\cite {rieger_phys-A_1994,rieger_prb_1996}. This growing length scale with waiting time in an out-of-equilibrium dynamics of glass-forming liquids~\cite{smarajit_rep-prog-phys_2016} has received less attention. In Ref.~\cite{parisi_1999}, the growth of the dynamic length scale as a function of waiting time is discussed. The length scale was extracted from a two-point spatial correlation function studied using Monte Carlo simulations. A Mode Coupling Theory (MCT) approach in the aging regime~\cite{saroj_prl_2012} shows that the nonstationary version of a three-point correlation function~\cite{biroli_prl_2006} can capture the volume of cooperatively rearranging regions (CRR), which grows with waiting time. In equilibrium dynamics, the existence of CRR is widely regarded as the primary cause of spatial correlations in inhomogeneous local dynamics, commonly referred to as dynamic heterogeneity (DH)~\cite{ediger_anu-rev-phys-chem_2000}. The corresponding dynamic length scale $\xi_d$ in equilibrium has been calculated from the finite-size scaling~\cite{smarajit_pnas_2009} and the block analysis~\cite{indra_prl_2017} of the four-point dynamic susceptibility $\chi_4$, spatial correlations of particle displacement $g_{uu}$~\cite{pole_phy-a_1998,indra_prr_2020}, etc.}

\SK{The rapid increase in relaxation time upon cooling a supercooled liquid is believed to be due to the growth of an amorphous order~\cite{smarajit_ann-rev_2014,takeshi_prl_2007,indra_prl_2018}, associated with a concomitant growth of a static length scale $\xi_s$, which has not yet been calculated in the aging regimes of a structural glass. However, Ref.~\cite{parisi_1999} mentioned that the waiting time dependence of a similar length scale can also be calculated from the radial distribution function. Methods like point-to-set correlations~\cite{Biroli_nat-phy_2008}, FSS of the minimum eigenvalue of Hessian~\cite{smarajit_phys-a_2012}, and FSS of $\alpha$-relaxation time~\cite{smarajit_pnas_2009,smarajit_pre_2012} are typically used to calculate $\xi_s$ in equilibrium dynamics.} 

\SK{Finite-size scaling (FSS)~\cite{binder_z-phys-b_1990}, involving the computation of various dynamical and static properties at different system sizes, has been well established~\cite{smarajit_pnas_2009,berthier_pre_2012,smarajit_pre_2012}.  In equilibrium dynamics, within a computationally accessible timescale, $\tau_\alpha$, the $\alpha$-relaxation time (see \textit{Appendix A}), grows inversely with system size~\cite{smarajit_pnas_2009,smarajit_pre_2012} in three dimensions. In Ref.~\cite{berthier_pre_2012}, ``Kac-Fredrickson-Andersen'' (KFA) lattice-gas model~\cite{andersen_prl_1984} was introduced, which showed that in the long-range interaction limit of the model, $\tau_\alpha$ can have a non-monotonic growth with system size. Despite that, simulations of model glass-forming liquids showed that less fragile models fail to exhibit non-monotonicity and more fragile models display only a weak non-monotonicity~\cite{berthier_pre_2012,berthier_phys-proc_2012}, particularly below the MCT temperature~\cite{Kirkpatrick_pra_1985}. With current computational power, equilibrium molecular dynamics for various system sizes at very low temperatures is challenging; at the same time, finite-size studies of aging dynamics are feasible. Aging dynamics can, in principle, provide access to deep glassy regimes across different waiting times and system sizes~\cite {parisi_1997}.}

\begin{figure*}[htp!]
    \centering
    \includegraphics[page=1, trim=0cm 4.2cm 0cm 0cm, clip, width=1.00\textwidth]{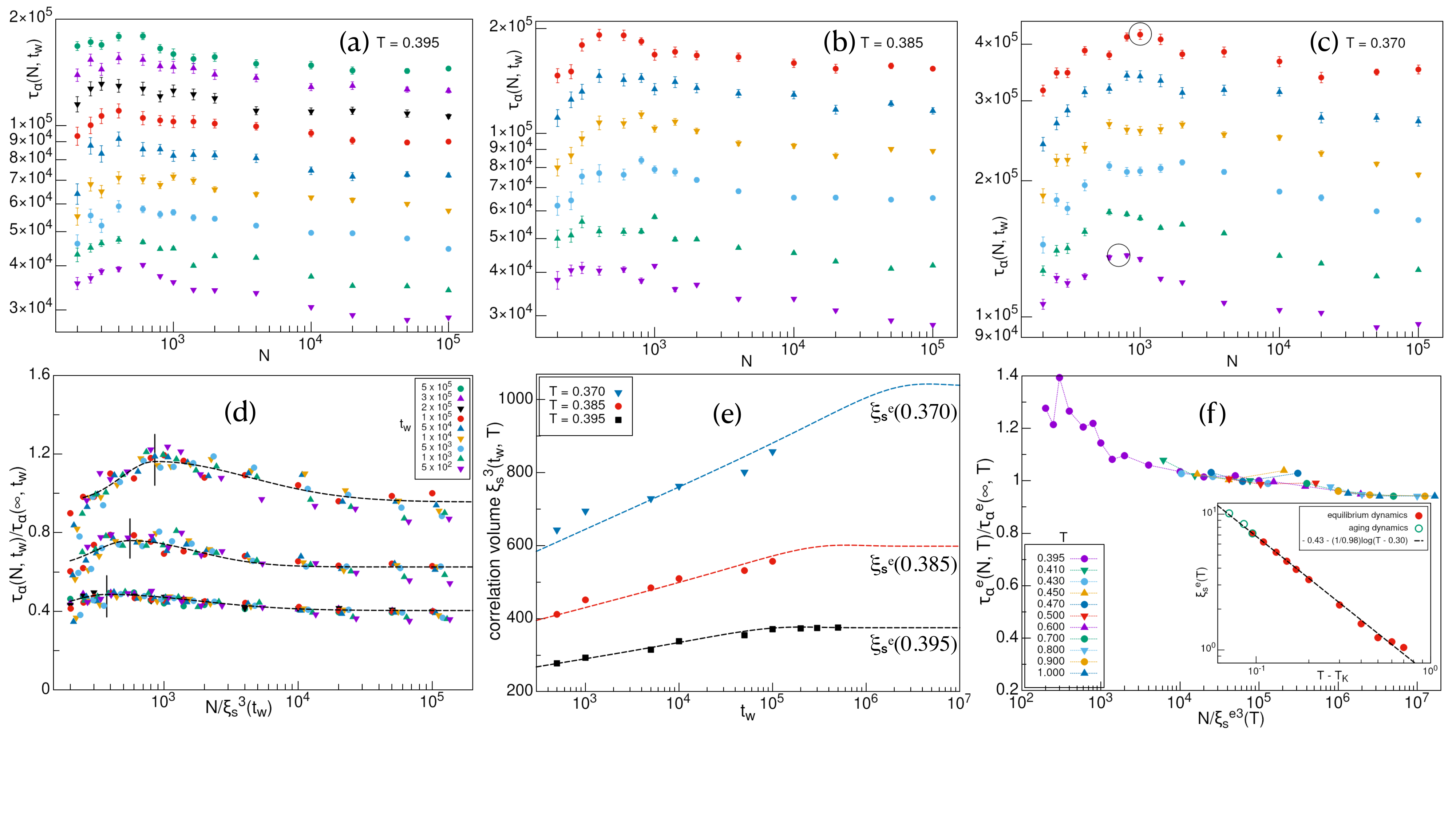}
    \caption{\SK{System size dependence of $\tau_\alpha(N,t_w)$ over different waiting times $t_w$ at temperatures (a) $T = 0.395$, (b) $T = 0.385$, and (c) $ T = 0.370$. The data corresponding to the maximum $t_w$ are plotted without rescaling, whereas the others are rescaled by an appropriate constant for clarity (see the SM~\cite{supplement} for the raw data). Black circles in (c) guide the eye, showing the apparent peak shift with $t_w$. (d) FSS in aging\textemdash Data collapse is obtained by rescaling the x and y-axis using $\xi_s^3(t_w)$ and $\tau_\alpha(\infty, t_w)$, respectively. Collapsed data are fitted with an asymmetric Gaussian function and rescaled by an appropriate constant. (e) Shows growth of the correlation volume $\xi_s^3(t_w, T)$ with $t_w$, calculated from peak positions in (d) for different $T$. The static length scale in equilibrium $\xi_s^{e}(T)$ is estimated by fitting $\xi_s^3(t_w, T)$ with Eq.~\ref{equ_proposed}. (f) FSS in equilibrium\textemdash Similar data collapse as (d) is shown for aging ($T = 0.395$) and equilibrated data ($T \ge T_g$)~\cite{kallol_2021} for $N \ge 4000$. The inset shows the estimated $\xi_s^e(T)$ from FSS in aging and equilibrium dynamics. The line is a power-law fit according to the RFOT theory.}}
    \label{figure_fss_tau}
\end{figure*}

\SK{In this letter, we present a detailed finite-size-scaling analysis of a model glass-forming liquid in an out-of-equilibrium aging state. We performed molecular dynamics simulations with measurements in different waiting times and across various system sizes. The static length scale associated with amorphous order is obtained from the FSS analysis of the dependence of $\tau_\alpha$ on system size. We show that the correlation volume grows logarithmically with waiting time in the aging regime, and the estimated size of this volume provides a reliable estimation of the correlation volume in equilibrium, which is consistent with the Random First Order Transition (RFOT) theory~\cite{kirkpatrick_pra_1989,lubchenko_ann-rev_2007,kirkpatrick_rmp_2015} prediction. The length scale associated with dynamic heterogeneity, obtained from the FSS and block analysis of $\chi_4$, is found to show a power-law dependence on waiting time.}

\vskip +0.05in
\noindent{\bf \large Simulation Details:} \SK{We have studied an $80:20$ binary mixture (A and B-type) of the Kob-Andersen model~\cite{kob_PRE_1995}, in three dimensions (3d). We simulate the system consisting of $N = 200$\textendash$100\,000$ particles, while keeping the number density fixed at $\rho = N/L^3 = 1.20$ ($L$ is the box length) with an integration time-step of $5 \times 10^{-3}$. Constant volume and temperature (NVT) molecular dynamics simulation has been performed using the Nosé–Hoover thermostat~\cite{nose_1984} by utilizing our in-house code and LAMMPS~\cite{plimpton_1995}. An out-of-equilibrium scenario has been generated by initially equilibrating at temperature $T_i = 5.00$ and then rapidly quenching to the final temperatures $T = 0.395, 0.385$, and $0.370$, below the glass-transition temperature ($T_g \sim 0.40$ for this model). All the measurements have been performed, after evolving at $T$, for a waiting time $t_w \in \{10^1, 5\times10^5\}$. Depending on the system size, $16$ to $768$ independent realizations were used to improve statistical accuracy. All relevant quantities reported here are computed for A-type particles unless otherwise mentioned. Further details of simulations can be found in the Supplementary Material (SM)~\cite{supplement}.}

\vskip +0.05in
\noindent{\bf \large FSS of $\alpha$-relaxation time:} \SK{We consider the dependence of the $\alpha$-relaxation time $\tau_\alpha(N, t_w)$ (defined in the \textit{Appendix A}) on the system size $N$ at the waiting time $t_w$, are shown in the top panel of Fig.~\ref{figure_fss_tau}(a\textendash c) for different temperatures. The data shows a non-monotonic growth of $\tau_\alpha(N, t_w)$ with increasing $N$, and saturates at a waiting time dependent value $\tau_\alpha(\infty,t_w)$. A careful observation also reveals that this growth is more pronounced as $T$ decreases, and the peak position shifts to larger system sizes with increasing $t_w$ at a fixed $T$. However, in equilibrium conditions,~\cite{smarajit_pnas_2009,smarajit_pre_2012} $\tau_\alpha$ decreases with increasing $N$ and saturates at larger $N$ for $T > T_g$, which also partially reflects in the aging regime after that non-monotonicity. In similarity with equilibrium dynamics, the dependence of $\tau_\alpha(N, t_w)$ on the system size $N$ is expected to exhibit the following finite-size scaling form:
\begin{equation}
    \label{equ_fss_tau}
    \tau_{\alpha}(N, t_w) = \tau_\alpha(\infty, t_w) \mathcal{F}\big(N/\xi_s^3(t_w)\big),
\end{equation}
where $\mathcal{F}(x)$ is an unknown scaling function, and $\xi_s(t_w)$ is the static length scale at $t_w$. The data for all waiting times at a fixed $T$ can be collapsed to a master curve using two parameters $\tau_\alpha(\infty, t_w)$ and $\xi_s(t_w)$, as shown in Fig.~\ref{figure_fss_tau}(d). The quality of the data collapse is satisfactory, and the peak position shifts to a larger rescaled value, $N/\xi_s^3(t_w)$, as the temperature decreases.} 

\SK{The observed non-monotonic growth of $\tau_\alpha(N, t_w)$ with system size $N$ at the waiting time $t_w$ is related to the growth of the static length scale $\xi_s(t_w)$ with waiting time.
In Ref.~\cite{berthier_pre_2012}, it was argued that an upper bound of relaxation time exists at equilibrium for a given system size $N$, as $\tau_\alpha^e(N,T) \sim \exp{(kN/T)}$, when the system size is smaller than the characteristic static length scale. Here $k$ is a constant and $T$ is the temperature. The argument is shown to be exact for a system with discrete degrees of freedom and evolving via stochastic dynamics~\cite{fabio_1999}. If we extend this argument for the aging system as a function of waiting time, then we would expect the relaxation time to grow with increasing $N$, and the growth will be stronger as $t_w$ increases due to the growth of the underlying length scale with $t_w$. In Fig.~\ref{figure_fss_tau}(a\textendash c), we see that clearly for various aging temperatures. A detailed numerical analysis of the KFA model~\cite{berthier_pre_2012}, where the relative importance of Mode Coupling dynamics and cooperative dynamics can be controlled, suggested that the relaxation time first grows with system size and then decreases at large system size with a peak at an intermediate system size. We see, probably for the first time, a clear indication of this behavior in our aging study of structural glass. A similar argument in Ref.~\cite{george_pre_2012} argued that, in equilibrium dynamics with cooperative relaxation process, one expects to have a system size dependence of $\tau_\alpha^e(N, T)$ that goes as $1/N$ for a large system size limit. Thus, the observed finite-size behavior of the relaxation time with waiting time in an aging glass-forming liquid indicates a very clear crossover from Mode Coupling dynamics to activated dynamics.}

\SK{In Fig.~\ref{figure_fss_tau}(e), we show the growth of the correlation volume $\xi_s^3(t_w, T)$ with waiting time for different $T$, which is extracted from the FSS of $\tau_\alpha(N,t_w)$ for various quenching temperature. Following Refs.~\cite{rieger_epl_1994,rieger_prb_1996}, we have fitted $\xi_s(t_w)$ in the aging regime using logarithmic fits $\xi_s(t_w) \sim \xi_{s_0} + [\ln(t_w)]^{1/\psi}$ with $\psi$ = $1.21, 1.52, 1.65$ which is in agreement with $\psi \le (d-1)$ in spin-glass ($d$ is the dimension)~\cite{fisher_prb_1988} and algebraic fits $\xi_s(t_w) \sim t_w^{\gamma}$ with $\gamma = 0.016, 0.017, 0.017$ for $T = 0.370, 0.385, 0.395$ respectively. The data for all these analyses are shown in detail in the SM~\cite{supplement}. Following Refs.~\cite {parisi_1997,parisi_1999}, we also computed the growth of $\xi_s(t_w)$ from the radial distribution function, which agrees well with FSS analysis as shown in \textit{Appendix B}. To estimate the static length scale at equilibrium $\xi_s^e(T)$ from the aging data in Fig.~\ref{figure_fss_tau}(e), we fitted $\xi_s^3(t_w, T)$ with the following equation,
\begin{equation}
    \label{equ_proposed}
    \small \xi_s^3(t_w, T) = {\xi_s^e}^3(T)\bigg[1 + \log\bigg(\frac{t_w}{\tau^e_\alpha(T)}\bigg)\exp\bigg(\frac{- \,\, t_w}{\tau^e_\alpha(T)}\bigg)\bigg] .
\end{equation}
Here $\tau^e_\alpha(T)$ is the $\alpha$-relaxation time in equilibrium at temperature $T$, computed on large $N$, taken from Ref.~\cite{pallabi_2022}. As shown in Fig.~\ref{figure_fss_tau}(e), the fit saturates at $t_w \geq \tau^e_\alpha(T)$, giving the value of $\xi_s^e(T)$ for different temperatures.}

\SK{Within our simulation timescale, the aging data of $\tau_\alpha(N, t_w)$ for $T = 0.395$ and $N \ge 4000$, can be considered as equilibrated data for $t_w \ge 2 \times 10^5$. Note that the smaller system size data is still in the aging regime. So, if we consider data at $t_w = 7 \times 10^5$ and $\tau^e_\alpha(N, T)$ in equilibrium dynamics for $T\ge T_g$ (taken from Ref.~\cite{kallol_2021}), we can perform a similar FSS at equilibrium as shown in Fig.~\ref{figure_fss_tau}(f). From this FSS analysis, we estimated the $\xi^e_s(T)$ for $T = 0.395$, so that we can directly compare the length scale from equilibrium data and the aging data. Within the RFOT theory~\cite{kirkpatrick_pra_1989,lubchenko_ann-rev_2007,kirkpatrick_rmp_2015}, $\xi^e_s(T)$ in equilibrium, should grow with $T$ as $\xi_s^e(T) \sim (T -T_K)^{-1/(d-\theta)}$ where, $\theta$ is the scaling exponent, and the Kauzmann temperature $T_K \simeq 0.30$ for this model. As shown in the inset of Fig.~\ref{figure_fss_tau}(f), the fitting parameter $(d - \theta) = 0.98$ or $\theta = 2.02$, nearly matches with the obtained values in experiment~\cite{biroli_bouchaud_2012} and simulation~\cite{kallol_2021}.}

\begin{figure*}[htp!]
    \centering
    \includegraphics[page=1, trim=0cm 3.8cm 0cm 0cm, clip, width=1.00\textwidth]{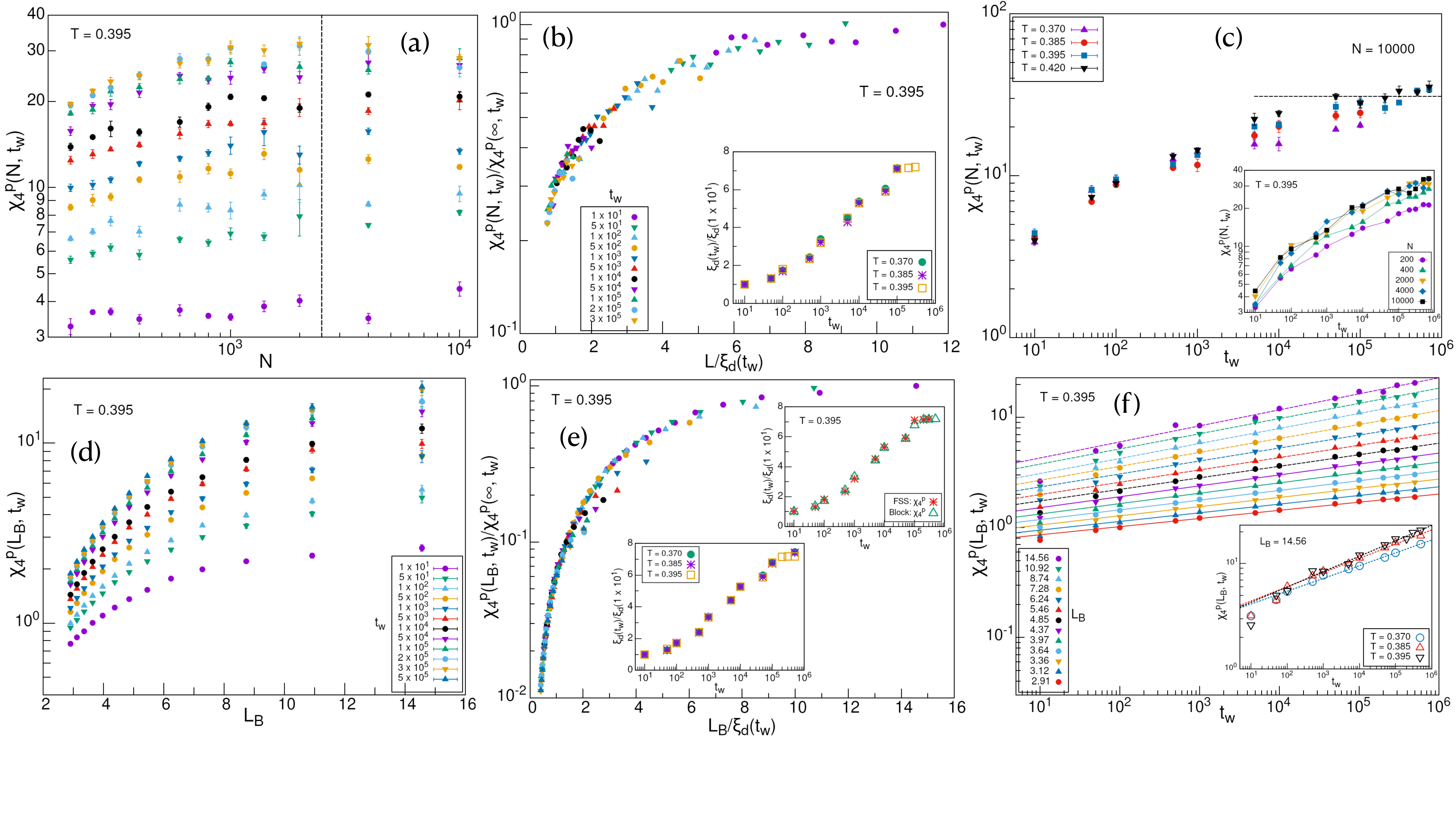}
    \caption{\SK{(a) System size dependence of $\chi_4^p(N, t_w)$ for different $t_w$ are shown for $T = 0.395$. The dotted line shows the system size at which $\chi_4^p$ becomes constant ($N\ge2000$). (b) A collapse of the data is obtained by rescaling x and y-axis by $\xi_d(t_w)$ and $\chi_4^p(\infty, t_w)$. The inset shows the growth of the normalised $\xi_d(t_w)$ for different temperatures, and saturates at longer $t_w$ for $T = 0.395$. (c) Waiting time dependence of $\chi_4^p$ for different $T$ is shown, and $\chi_4^p$ seems to saturate at longer $t_w$ as it approaches the equilibrium. Inset shows $t_w$ dependence of $\chi_4^p$ for different system sizes at $T=0.395$. (d) Block size dependence of $\chi_4^p$ for different $t_w$ is shown at $T = 0.395$ and $N = 100\,000$. (e) Similar data collapse as (b) is also obtained in block analysis. Lower inset shows growth of $\xi_d(t_w)$ with $t_w$ for different $T$. The upper inset shows a comparison of $\xi_d(t_w)$ with $t_w$ from FSS and block analysis at $T = 0.395$. (f) Waiting time dependency of $\chi_4^p$ for different block sizes at $T = 0.395$ and different $T$ at $L_B = L/3$ is shown in the main and inset, respectively.}}
    \label{figure_fss_chi4}
\end{figure*}

\noindent{\bf \large FSS of Dynamical Heterogeneity:} \SK{We consider the system size dependence of the peak value of $\chi_4(N, t_w)$ (defined in \textit{Appendix C}), denoted as $\chi_4^p(N, t_w)$, as shown in Fig.~\ref{figure_fss_chi4}(a) for different waiting times at $T = 0.395$. Similar to equilibrium scenario, $\chi_4^p(N, t_w)$ in aging conditions at a fixed $t_w$, also grows with $N$ and saturates at $\chi_4^p(\infty, t_w)$. 
Likewise, in the block analysis~\cite{indra_prl_2017}, we investigate the block size $L_B$ dependence on the peak value of $\chi_4(L_B, t_w)$ (defined in \textit{Appendix C}) for different waiting times, as shown in Fig.~\ref{figure_fss_chi4}(d). These results are qualitatively similar. The $\chi_4^p(L_B, t_w)$ with variation of $L_B$ and $t_w$, is performed on a relatively larger system of size $N = 100\,000$. According to the FSS hypothesis~\cite{binder_z-phys-b_1990}, the system size dependence of $\chi_4^p(N, t_w)$ is expected to have the form
\begin{equation}
    \label{equ_fss_chi4}
    \chi_4^p(N, t_w) = \chi_4^p(\infty, t_w) \mathcal{G}\big(L/\xi_d(t_w)\big),
\end{equation}
where $\mathcal{G}(x)$ is an unknown scaling function, and $\xi_d(t_w)$ is the dynamic length scale at $t_w$. The data of $\chi_4^p$ at a fixed $T$, collapses onto a single scaling curve using $\chi_4^p(\infty, t_w)$ and $\xi_d(t_w)$, as shown in Fig.~\ref{figure_fss_chi4}(b) and  Fig~\ref{figure_fss_chi4}(e). The scaling collapse of $\chi_4^p$ in Fig.~\ref{figure_fss_chi4}(e) is better than that in Fig.~\ref{figure_fss_chi4}(b), since block analysis involves averaging over both different blocks and realizations, whereas in the latter case the averaging is only over different realizations.} 

\SK{From Eq.~\ref{equ_fss_chi4}, if we assume that $\chi_4^p(\infty, t_w) \propto \xi_d^{2-\eta}$, then one would expect $\mathcal{G}(x) \propto x^{2-\eta}$ for $x \rightarrow 0$. This will suggest that $\chi_4^p(N, t_w) \propto L^{2-\eta}$ in the small $L/\xi_d(t_w)$ limit. We have estimated the exponent $\eta$ from the collapsed data in block analysis method and found $\eta \simeq -0.23$ (see Supplementary Material~\cite{supplement}), which is consistent with earlier results~\cite{indra_prl_2017,smarajit_prl_2010} in equilibrium dynamics. The growth of $\xi_d(t_w)$ with $t_w$ for different temperatures is obtained from the FSS and block analysis of $\chi_4$ is shown in the inset of Figs.~\ref{figure_fss_chi4}(b) and \ref{figure_fss_chi4}(e), respectively. This growth follows a power law: $\xi_d(t_w) \sim t_w^\gamma$, with the exponent $\gamma$ ranging between $0.184\textendash0.194$ in both FSS and block analysis across different temperatures, in excellent agreement with the predicted value of $1/6$ in structural glass~\cite{parisi_1997,parisi_1999}. A detailed description of these analyses is reported in Supplementary Material~\cite{supplement}. The obtained values of $\xi_d$ from FSS and block analysis are well in agreement with each other, shown for $T = 0.395$ in the upper inset of Fig.~\ref{figure_fss_chi4}(e).}

\begin{figure*}[htp!]
    \centering
    \includegraphics[page=1, trim=0cm 21cm 0cm 0cm, clip, width=1.00\textwidth]{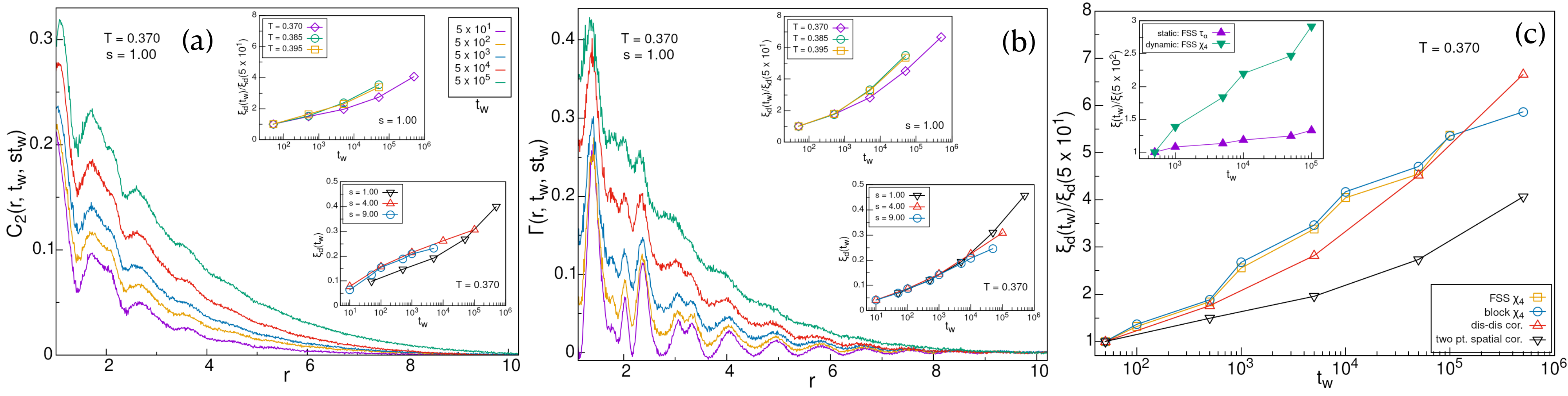}
    \caption{\SK{A comparison between (a) the two-point spatial correlation $C_2(r, t_w, st_w)$ and (b) the spatial displacement correlation $\Gamma(r, t_w, st_w)$ is shown for different $t_w$ at $T = 0.370$, $N = 100\,000$, and $s = 1.00$ (see text for details). Lower insets in (a) and (b) show the integrated area (proportional to $\xi_d(t_w)$) of $C_2(r, t_w, st_w)$ and $\Gamma(r, t_w, st_w)$ for different $s$ at $T = 0.370$. Similarly, the upper inset shows the normalised $\xi_d(t_w)$ for different temperatures at $s = 1.00$. (c) A qualitative comparison of the normalised $\xi_d(t_w)$ obtained from different methods is shown. The inset shows the growth of the normalised static and dynamic length scale with waiting time.}}
    \label{figure_scf}
\end{figure*}

\SK{It is interesting to observe that for lower temperatures, $\chi_4^p(N, t_w)$ is smaller than higher temperatures for a given $t_w$ (see Fig.~\ref{figure_fss_chi4}(c)), as it follows the envelope of the equilibrium $\chi_4(t)$ and for higher temperature the $\chi_4(t_w, t)$ will have larger value for a given waiting time as it peaks earlier (see the SM~\cite{supplement}). It corroborates well with the observation that $\chi_4^p(t)$ saturates or becomes smaller below MCT transition temperature~\cite{coslovich_2018, pallabi_2022}. Although it is not clear whether $\chi_4^p(N, t_w \ge \tau^e_\alpha(T))$ will be larger at lower temperature or not. Inset of Fig.~\ref{figure_fss_chi4}(c) shows that the variation of $\chi_4^p$ does not change significantly for $N\ge 2000$, also reported in Ref.~\cite{coslovich_2018}. In Fig.~\ref{figure_fss_chi4}(f), we show similar results obtained using the block analysis method.}

\vskip +0.05in
\noindent{\bf \large Dynamic Length Scale from Spatial Correlation Functions:} \SK{The two-point spatial correlation function~\cite{parisi_1999} and the spatial displacement correlation function~\cite{pole_phy-a_1998} (both defined in \textit{Appendix D}) have been computed in our analysis, and a comparison between them is shown in Fig.~\ref{figure_scf}. As waiting time increases, both functions exhibit progressively slower spatial decay, indicating that $\xi_d(t_w)$ increases with $t_w$. The integrated area of these spatial correlation functions gives the heterogeneity length scale, as shown in the lower inset of Fig.~\ref{figure_scf}(a) and \ref{figure_scf}(b). It turns out that for different values of $s$ (see \textit{Appendix D}), the obtained values of $\xi_d$ are almost the same, as previously observed in Ref.~\cite{parisi_1999}. In the upper inset, the growth of normalized $\xi_d(t_w)$ for different temperatures at $s = 1.00$ is shown. The obtained values of normalized $\xi_d(t_w)$ do not vary significantly with temperature, as the simulated temperatures are very close to each other. Algebraic fits yield $\gamma \approx 0.15\textendash0.18$ from two-point spatial correlations and $0.19\textendash0.24$ from spatial displacement correlations, in qualitative agreement with FSS and block analysis. Figure~\ref{figure_scf}(c) compares the obtained values of normalized $\xi_d(t_w)$ from different methods, showing good agreement among them. The inset shows the growth of normalized correlation lengths $\xi_d(t_w)$ and $\xi_s(t_w)$ with waiting time, consistent with the equilibrium scenario in which the static length scale is smaller than the dynamic length scale~\cite{smarajit_pnas_2009,smarajit_ann-rev_2014}.}

\vskip +0.05in
\noindent{\bf \large Conclusions:} \SK{In this letter, we performed finite-size scaling analysis in the aging regime of a glass-forming liquid, demonstrating that a logarithmic growth with aging time best describes the time evolution of the static length scale, while the dynamic length scale follows a power-law growth. We show how aging results can be used to estimate the growth of the static length scale in equilibrium, and this estimate agrees well with the predictions of the RFOT theory. The correlation lengths computed from spatial correlation functions are consistent with the finite-size scaling analysis, suggesting that aging dynamics are governed by growing static and dynamic length scales. As aging measurements in microgels and synthetic clays are comparatively easy to perform experimentally, we hope that this work will motivate experimental studies to measure length scales from the waiting-time description of the FSS and spatial correlation functions.} 

\vskip +0.05in
\begin{acknowledgments}
\noindent{\bf \large Acknowledgments:} We acknowledge the funding by intramural funds at TIFR Hyderabad from the Department of Atomic Energy (DAE) under Project Identification No. RTI 4007. SK would like to acknowledge Swarna Jayanti Fellowship Grant Nos. DST/SJF/PSA01/2018-19 and SB/SFJ/2019-20/05 from the Science and Engineering Research Board (SERB) and Department of Science and Technology (DST). SK also acknowledges research support from MATRICES Grant MTR/2023/000079 from SERB.
\end{acknowledgments}

\bibliographystyle{unsrt}
\bibliography{bibliog_main.bib}

\clearpage
\section{End Matter}
\begin{figure*}[htp!]
    \centering
    \includegraphics[page=1, trim=0cm 21.5cm 0cm 0cm, clip, width=1.00\textwidth]{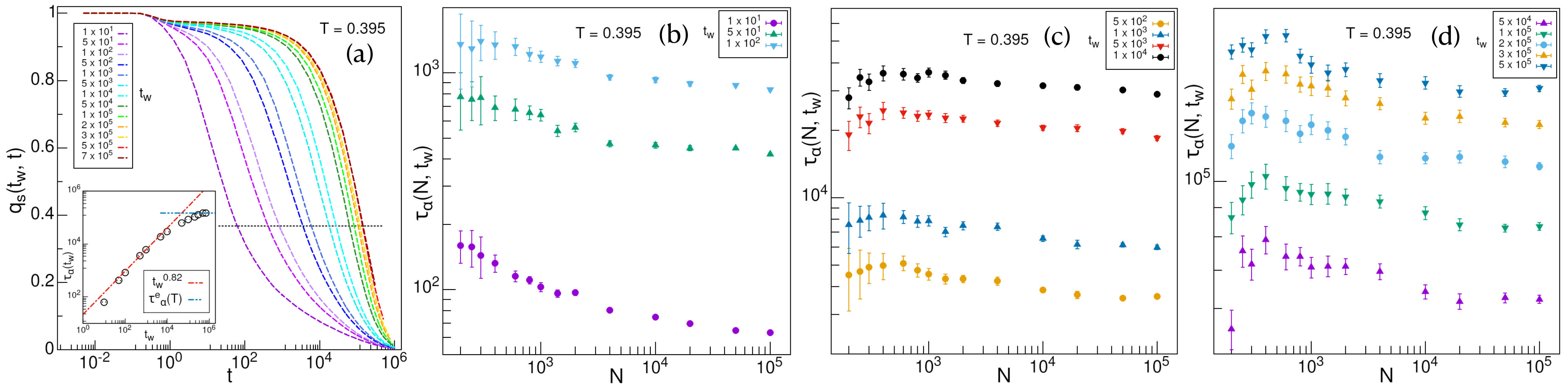}
    \caption{(a) Waiting time dependence of the self-overlap function is shown for $N = 100\,000$ at $T = 0.395$. The black dotted line shows the $1/e$ value, that used to calculate the $\alpha$-relaxation time of the corresponding $t_w$. Inset shows the growth of $\tau_\alpha$ with waiting time, which is fitted with $t_w^{0.82}$ in the aging regime and saturates for $t_w\ge \tau^e_\alpha(T)$ in equilibrium. System size dependence of $\tau_\alpha$ for different waiting times at $T = 0.395$ are shown in (b), (c), and (d). These data are rescaled using the scaled factors of $7.90, 5.70, 2.42, 1.98, 1.16, 1.08, 1.00, 1.00, 1.00$ for $t_w = 5\times10^2, 1\times10^3, 5\times10^3, 1\times10^4, 5\times10^4, 1\times10^5, 2\times10^5, 3\times10^5, 5\times10^5$ respectively, are shown in Fig.~\ref{figure_fss_tau} of the main text.}
    \label{figure_qoft}
\end{figure*}

\noindent{\bf Appendix A: Non-stationary variant of the overlap function\textemdash} The waiting time dependence of the self-overlap function~\cite{parisi_1997,rituparno_prl_2020} in an out-of-equilibrium dynamics is defined as
\begin{equation}
   \label{equ_qoft}
   q_s(t_w, t) = \frac{1}{N} \sum_{i=1}^{N} w\Big(\big|\Vec{r_i}(t_w + t) - \Vec{r_i}(t_w)\big|\Big),
\end{equation}
where the term $\big|\Vec{r_i}(t_w + t) - \Vec{r_i}(t_w)\big|$ denotes displacement of $i$-th particle over the time interval $t$, measured from the waiting time $t_w$, and the window function defined as $w(x) = \Theta(a -x)$, with $\Theta$ denoting the Heaviside step function. The parameter ``\textit{a}'' is chosen from the plateau height of the mean-squared displacement (MSD) in supercooled liquids, to remove the de-correlation arising from vibrations of particles inside the cages formed by their nearest neighbors. In our simulation model, we choose \textit{a} to be 0.30$\sigma_{AA}$, where $\sigma_{AA}$ is the diameter of the larger size particle. The $\alpha$-relaxation time for different waiting times is defined as $[q_s(t_w,t=\tau_\alpha)] = 1/e$, where $[...]$ denotes averaging over multiple realizations. As depicted in Fig.~\ref{figure_qoft}, the system size dependence of $\tau_\alpha(N, t_w)$ shows that the peak position is shifted to larger system size with increasing $t_w$ and saturates at $t_w \ge \tau^e_\alpha(T)$ in equilibrium. It signifies how the amorphous order (static length scale) is growing with waiting time, which is discussed in more detail in the main text.

\vskip +0.05in

\begin{figure}[htp!]
    \centering
    \includegraphics[keepaspectratio, width=0.48\textwidth]{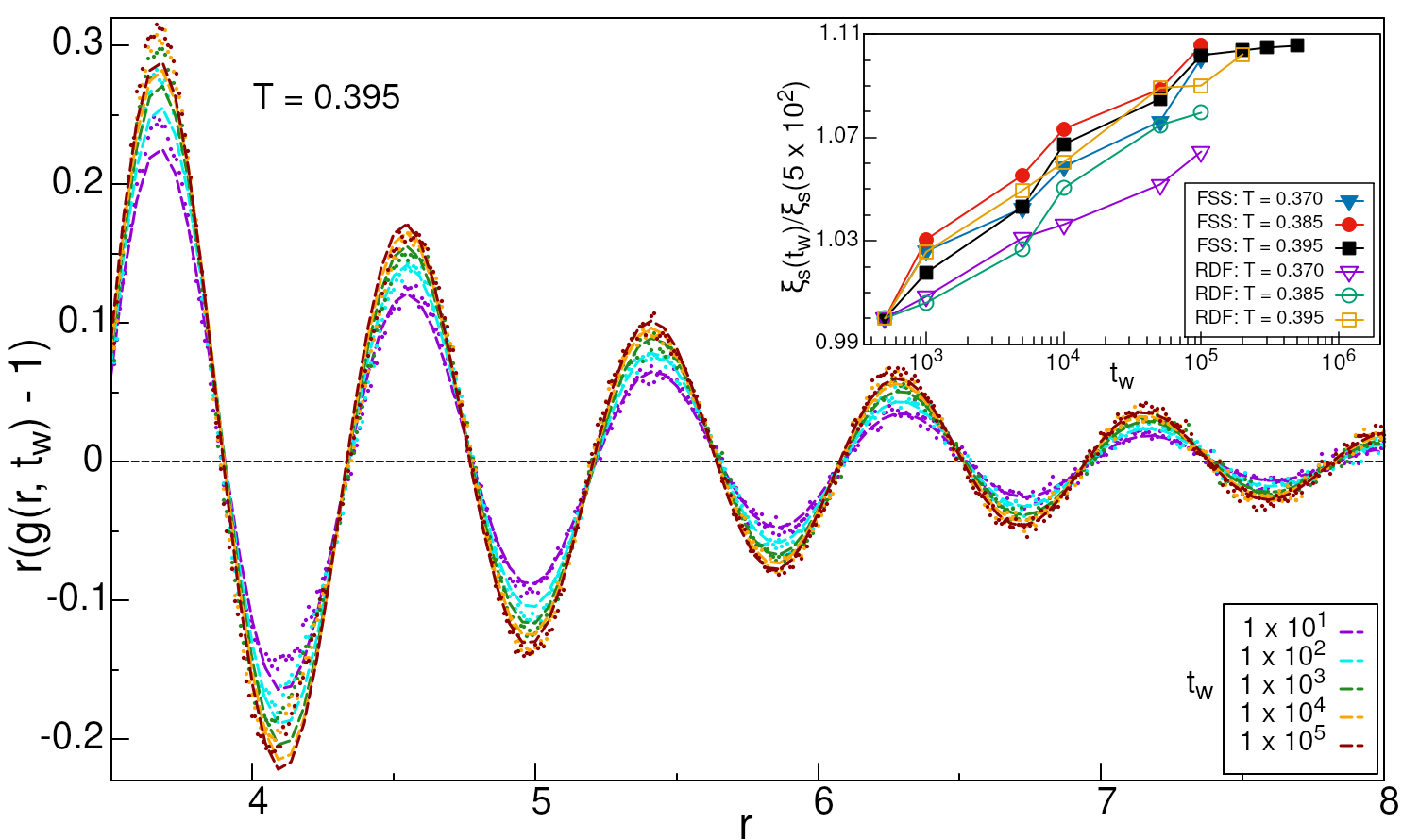}
    \caption{Waiting time variation of $r(g(r,t_w) - 1)$, fitted with Eq.~\ref{equ_gofr_fit}, is plotted against distance $r$ for $N = 100\,000$ at $T = 0.395$. Inset shows the comparison of normalized static length scales obtained from FSS of $\tau_\alpha$ and RDF.}
    \label{figure_gofr}
\end{figure}

\noindent{\bf Appendix B: Static length scale from radial distribution function\textemdash} The non-stationary variant of the radial distribution function (RDF)~\cite{parisi_1997,parisi_1999} computed at a distance $r$, from a reference particle at a waiting time $t_w$ is defined as
\begin{equation}
    \label{equ_gofr}
    g(r, t_w) = \frac{1}{4 \pi r^2 \Delta r \rho N} \sum_{i \ne j} \delta \Big(r - \big|\Vec{r_j}(t_w) - \Vec{r_i}(t_w)\big|\Big),
\end{equation}
where the term $\big|\Vec{r_j}(t_w) - \Vec{r_i}(t_w)\big|$ denotes the distance between $i$-th and $j$-th particle at $t_w$, $\delta(x)$ is a Dirac delta function, and $\Delta r$ denotes the radial interval. FSS of $\alpha$-relaxation time requires a systematic analysis over multiple system sizes, whereas a similar length scale can be extracted from RDF on a single, sufficiently large system size. As shown in Fig.~\ref{figure_gofr}, the function $r\big(g(r,t_w) - 1\big)$ fluctuates around zero, and amplitude increases with $t_w$. Following the Ref.~\cite{parisi_1999}, we extract $\xi_s(t_w)$ from $r\big(g(r,t_w) - 1\big)$ by fitting with the equation shown below,
\begin{equation}
    \label{equ_gofr_fit}
    r\big(g(r,t_w) - 1\big) \approx Br^{(1-c)} \exp{(-r/\xi_s(t_w))} \sin{(2\pi r/f + \phi)}.
\end{equation} 
The static length scale $\xi_s(t_w)$ is extracted by fitting the data for $r \ge 3.50\sigma_{AA}$, setting $c = 1$, and treating $B$, $f$, and $\phi$ as fitting parameters. Inset of Fig.~\ref{figure_gofr} shows the growth of normalized $\xi_s(t_w)$ obtained from RDF and FSS, where logarithmic fits yield $\psi = 1.24$\textendash$1.87$, demonstrating qualitative agreement between the two methods.

\begin{figure}[htp!]
    \centering
    \includegraphics[page=1, trim=0cm 5.3cm 0cm 0cm, clip, width=0.48\textwidth]{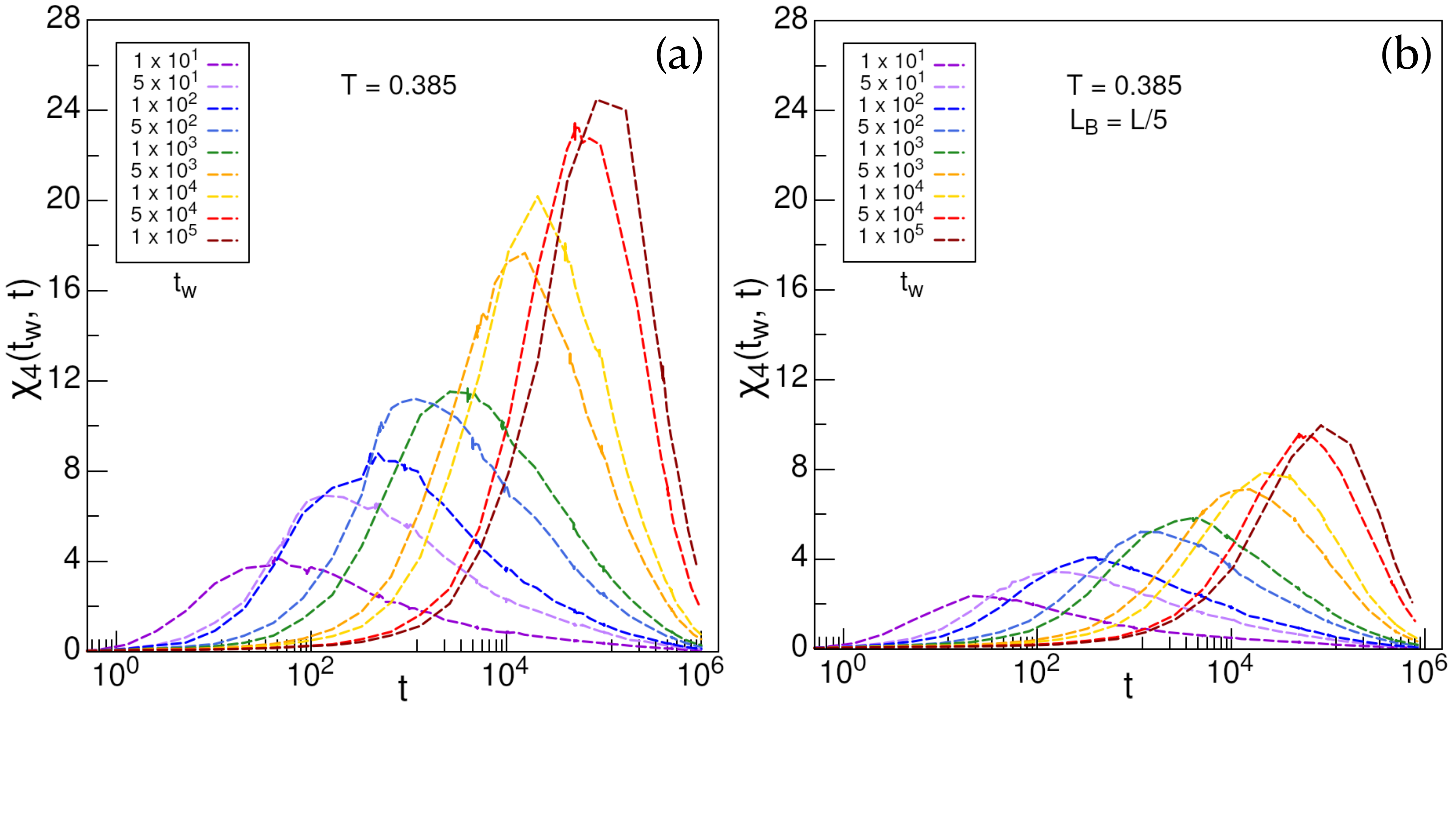}
    \caption{Waiting time dependence of $\chi_4$ is shown at $T = 0.385$ (a) without block analysis for $N =10\,000$, (b) with block analysis ($L_B = L/5$) for $N =100\,000$.}
    \label{figure_chi4}
\end{figure}

\vskip +0.05in

\noindent{\bf Appendix C: Dynamic susceptibility $\chi_4$ in aging dynamics\textemdash} Introduced by Dasgupta \textit{et. al.}~\cite{Dasgupta_1991}, the four-point dynamic susceptibility measures the dynamic heterogeneity~\cite{ediger_anu-rev-phys-chem_2000} of the system. The non-stationary version of $\chi_4$~\cite{parisi_1997,parisi_1999}, calculated from fluctuations of the two-point overlap function $q_s(t_w, t)$ across different realizations, is defined as
\begin{equation}
    \chi_4(t_w,t) = N \Big[q_s^2(t_w,t)\Big] - N \Big[ q_s(t_w,t) \Big]^2.
\end{equation}
In Fig.~\ref{figure_chi4}(a), we show the waiting time dependence of $\chi_4$ in the out-of-equilibrium dynamics. The dynamic length scale $\xi_d$, related to the heterogeneity of the system, is calculated from the FSS of $\chi_4$, is discussed in the main text. Another efficient and experimentally realizable method to obtain the waiting time variation of $\xi_d$, we performed ``block analysis"~\cite{indra_prl_2017} within FSS of $\chi_4$, where $\chi_4$ is computed over smaller subsystems of varying sizes within a large system. The size of the smaller subsystems (blocks) is given by $L_B = L/n$, where $n \in \{3, 4, 5 ... \}$. The self-overlap function and the corresponding dynamic susceptibility, associated with blocks of size $L_B$ at waiting time $t_w$, are defined as
\begin{align}
    \notag
    & q_s(L_B, t_w, t) = \\
    & \hspace{1cm} \frac{1}{N_B} \sum_{i=1}^{N_B} \frac{1}{n_i(t_w)} \sum_{j=1}^{n_i(t_w)} w\Big(\big|\Vec{r_j}(t_w + t) - \Vec{r_j}(t_w)\big|\Big), \\
    \notag
    & \chi_4(L_B,t_w,t) = \\
    & \hspace{1cm}  \frac{N}{N_B}  \left[\Big<q_s^2(L_B,t_w,t)\Big> - \Big< q_s(L_B,t_w,t) \Big>^2 \right],
\end{align}
where, $N_B$ is the number of blocks with size $L_B$, $n_i(t_w)$ is the number of particles in the $i$-th block at $t_w$. $\left<...\right>$ and $[...]$ denote averaging over different blocks of size $L_B$, and different realizations, respectively. In Fig.~\ref{figure_chi4}(b) we show the variation $\chi_4$ with $t_w$ from block analysis, with size of block $L_B = L/5$. A comparison between Fig.~\ref{figure_chi4}(a) and \ref{figure_chi4}(b) reveals that the height of $\chi_4$ decreases and appears smoother in the block analysis. 

\vskip +0.05in

\noindent{\bf Appendix D: Spatial correlation functions in out-of-equilibrium dynamics\textemdash} To calculate age dependent dynamic length scale $\xi_d$, Parisi introduced a two-point spatial correlation function in Ref.~\cite{parisi_1999}, which has a close similarity to the usual ferromagnetic correlation function~\cite{huang_stat_book}. The full overlap function for each particle $q_i(t_w,t)$, and its average value $q(t_w,t)$ at waiting time $t_w$ and time interval $t$, are given by
\begin{align}
    q_{i}(t_w,t) &= \sum_{j=1}^N w\Big(\big|\Vec{r}_i(t_w+t) - \Vec{r}_j(t_w)\big|\Big),\\
    q(t_w,t) &= \frac{1}{N}\sum_{i=1}^N q_{i}(t_w,t),
\end{align}
where the term $|\Vec{r}_i(t_w+t) - \Vec{r}_j(t_w)|$ is defined as the distance between $i$-th particle at time $(t_w+t)$ and $j$-th particle at the waiting time $t_w$. By utilizing the full overlap function $q_{i}(t_w,t)$, a spin-like variable is introduced as $\sigma_i(t_w,t)= 2(q_{i}(t_w,t) -  q(t_w,t))$. For $q(t_w, t)$ close to $\frac{1}{2}$, the spin-like variable takes the value $+1$, if a particle moves less than $a$ (defined earlier) and $-1$, if it moves more than $a$. The sum of this spin-like variable over all particles is zero, resembling a state of zero net magnetization. The two-point spatial correlation function at a distance $r$ from a reference particle is given by
\begin{align}
    \small f(r,t_w,t) = \Bigg[\frac{\sum_{i \ne j} \sigma_i(t_w,t) \sigma_j(t_w,t) \delta\big(r - |\Vec{r_{ij}}(t_w)|\big)}{4\pi r^2 \Delta r N \rho}\Bigg],
\end{align}
where $\Delta r$ denotes the radial interval and the term $\Vec{r_{ij}}(t_w) = \vec{r}_j(t_w) - \vec{r}_i(t_w)$ defines the relative position between $i$-th and $j$-th particle at $t_w$. To eliminate the natural oscillatory effects, we compute the excess two-point spatial correlation function $C_2(r,t_w,t) = f(r,t_w,t)\big/g(r)$, where $g(r)$ denotes the usual radial distribution function, averaged over multiple time origins. The time difference is chosen as $t = s\,t_w$, where ``$s$'' is a scaled multiple of waiting time $t_w$.

The waiting time dependence of the spatial correlation function calculated from particle displacements~\cite{pole_phy-a_1998} captures the growth of $\xi_d$ more effectively than the two-point spatial correlation function, which is represented as
\begin{equation}
    g_{uu}(r, t_w, t) =  \frac{\Big[\sum_{i \ne j} u_i(t_w, t) \, u_j(t_w, t) \,\, \delta (r - |\vec{r_{ij}}(t_w)|) \Big]}{4\pi r^2 \Delta r N \rho \, \Big[ u(t_w, t)\Big]^2},
\end{equation}
where $u_i(t_w,t) = |\vec{r_i}(t_w+t) - \vec{r_i}(t_w)|$ is the displacement of $i$-th particle between times $t_w$ and $t_w +t$. The term $u(t_w,t) = \frac{1}{N} \sum_{i=1}^N u_i(t_w,t)$ is the average displacement of particles in time difference $t$, measured from the waiting time $t_w$. To extract the dynamic length scale $\xi_d$, we calculate the excess spatial displacement correlation $\Gamma(r, t_w, t) = \big(g_{uu}(r, t_w, t)\big/g(r)\big) - 1$, where the time difference is taken as $t = st_w$.
\end{document}


\title{Growth of Dynamic and Static Correlations in the Aging Dynamics of a Glass-Forming Liquid: Supplementary Material}

\author{Santu Nath}
\email{snath@tifrh.res.in}
\affiliation{Tata Institute of Fundamental Research, 36/P, Gopanpally Village, Serilingampally Mandal, Ranga Reddy District, Hyderabad 500046, Telangana, India}

\author{Smarajit Karmakar}
\email{smarajit@tifrh.res.in}
\affiliation{Tata Institute of Fundamental Research, 36/P, Gopanpally Village, Serilingampally Mandal, Ranga Reddy District, Hyderabad 500046, Telangana, India}

\maketitle
\section*{S1. Model Details and Simulation Protocols}
We have studied the well-known Kob-Anderson binary mixture of model glass-forming liquid in three dimensions (3DKA model). The mixture ratio of the two types of particles (A and B types) is $80:20$. Particles are interacting via the Lennard-Jones potential, including the correction terms, such that the potential and force go to zero at the cutoff:
\[
V_{\alpha \beta}(r) =
\begin{cases}
         4\epsilon_{\alpha \beta} \left[\left(\frac{\sigma_{\alpha \beta}}{r}\right)^{12} - \left(\frac{\sigma_{\alpha \beta}}{r}\right)^{6} + c_0 + c_2 \left(\frac{r}{\sigma_{\alpha \beta}}\right)^2 \right], \hspace{0.5cm}\text{for } r < r_c \sigma_{\alpha \beta}, \\
         0, \hspace{6.9cm} \text{for } r \ge r_c \sigma_{\alpha \beta},
\end{cases}
\]
where $\alpha$ and $\beta$ signify types of interacting particles (AA, AB, BB). The cutoff distance $r_c$ is fixed at $2.50 \sigma_{AA}$. Diameter and energy parameters are $\sigma_{AB}$ = 0.80$\sigma_{AA}$, $\sigma_{BB}$ = 0.88$\sigma_{AA}$, and $\epsilon_{AB}$ = 1.50$\epsilon_{AA}$, $\epsilon_{BB}$ = 0.50$\epsilon_{AA}$, respectively. In this 3DKA model, length, energy, and time units are given by $\sigma_{AA}$, $\epsilon_{AA}$, and \( \sqrt{\sigma_{AA}^2 / \epsilon_{AA}} \) respectively. Masses of each type of particle and the Boltzmann constant are chosen as $m_A$ = $m_B$ = 1, and $K_B = 1$, respectively. The number density ($\rho = N/V$, where $N$ is the system size and $V$ is the volume of the simulation box) is fixed at $1.20\sigma_{AA}^{-3}$. To study finite-size scaling in out-of-equilibrium dynamics, we have performed extensive molecular dynamics (MD) simulations using an in-house code for $N = 200$ to $2000$, in 9 different system sizes, and using LAMMPS~\cite{lammps} for $N = 4000$ to $100\,000$, in 5 different system sizes. We have performed NVT simulation using the \textit{Nosé–Hoover}~\cite{nose_1984} thermostat to maintain a constant temperature, with an integration time step of $dt = 0.001$ at higher temperatures, and $dt = 0.005$ at lower temperatures.\\

To create an out-of-equilibrium scenario, simulations have been performed in the following way: starting from an equilibrium configuration at high temperature ($T_i = 5.00$), we quenched the system at time $t = 0$ to a final temperature $T$. In this 3DKA model, the calorimetric glass-transition temperature is around $T_g \sim 0.40$, and the final temperatures are fixed at $T  = 0.395$ ($1.25\%$ below $T_g$), $0.385$ ($3.75\%$ below $T_g$), and $0.370$ ($7.50\%$ below $T_g$). The corresponding $\alpha$-relaxation times $\tau^e_\alpha(T)$ at equilibrium span a broad range from $2 \times 10^5$ to $6 \times 10^6$ for large system size, depending on the final temperature $T$~\cite{pallabi_2022}. Before taking data, each system is evolved up to a waiting time $t_w$ ranging from $1 \times 10^1$ to $7 \times 10^5$, ensuring that the system remains out of equilibrium within the total simulation timescale of $1 \times 10^6$. Analysis of time correlation functions in out-of-equilibrium dynamics needs a strong averaging for finite-size scaling analysis. We have taken different ranges of statistically independent initial configurations for various system sizes, which are listed below in Table~\ref{table_mdruns}. In this work, all types of correlation functions are computed for A-type particles unless stated otherwise.
\begin{table}[htp]
    \centering
    \begin{tabular}{|c|c|c|c|c|c|c|c|c|}
    \hline
    System size (N) & 200 to 800 & 1000 & 1400 to 10\,000 & 20\,000 to 50\,000 & 100\,000 \\
    \hline
    Number of realizations & 768 & 512 & 384 & 32 & 16\\
    \hline
    \end{tabular}
    \caption{Variation in the number of realizations used for statistical averaging for different system sizes.}
    \label{table_mdruns}
\end{table}

\clearpage
\section*{S2. Additional Figures}
\section*{A. FSS of $\alpha$-relaxation time}
\begin{figure*}[htp!]
    \centering
    \includegraphics[page=1, trim=0cm 15.8cm 0cm 0cm, clip, width=1.00\textwidth]{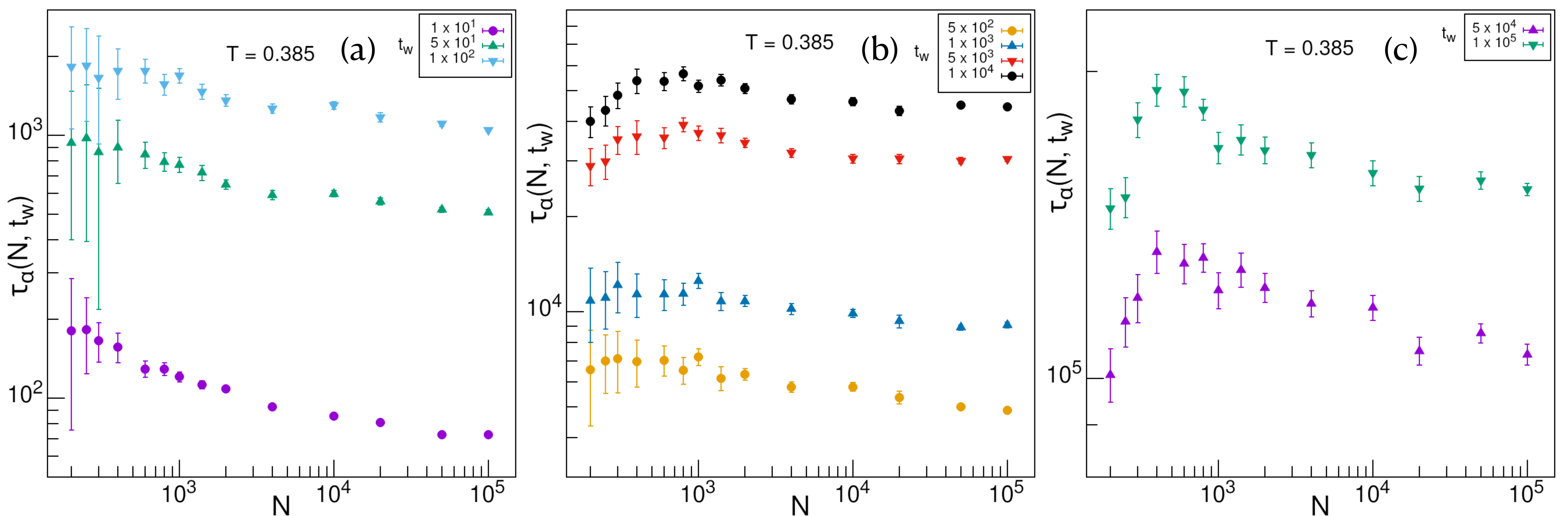}
    \caption{System size dependence of the $\alpha$-relaxation time for different waiting times at $T = 0.385$ are shown in (a), (b), and (c). These data are rescaled using the scaled factors of $5.80, 4.60, 2.15, 2.00, 1.10, 1.00$ for $t_w = 5\times10^2, 1\times10^3, 5\times10^3, 1\times10^4, 5\times10^4, 1\times10^5$ respectively, are shown in the main text.}
    \label{tau_vs_N_T0.385}
\end{figure*}

\begin{figure*}[htp!]
    \centering
    \includegraphics[page=1, trim=0cm 15.8cm 0cm 0cm, clip, width=1.00\textwidth]{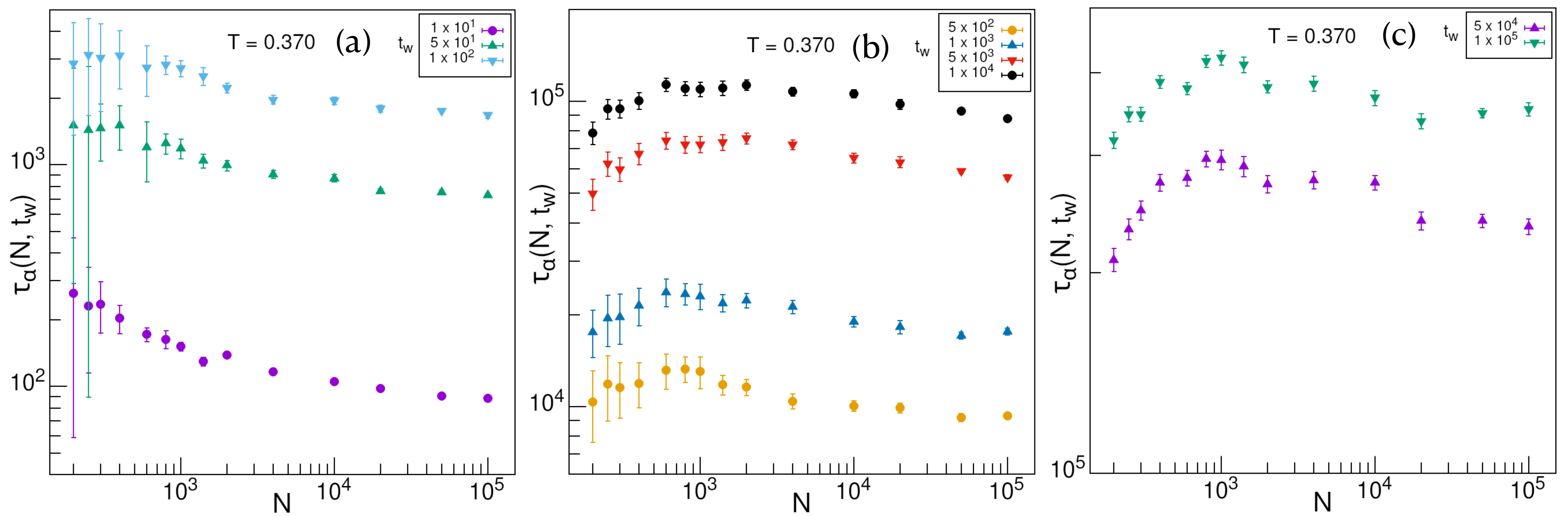}
    \caption{System size dependence of the $\alpha$-relaxation time for different waiting times at $T = 0.370$ are shown in (a), (b), and (c). These data are rescaled using the scaled factors of $10.30, 7.20, 2.90, 2.35, 1.15, 1.00$ for $t_w = 5\times10^2, 1\times10^3, 5\times10^3, 1\times10^4, 5\times10^4, 1\times10^5$ respectively, are shown in the main text.}
    \label{tau_vs_N_T0.370}
\end{figure*}

\begin{figure*}[htp!]
    \centering
    \includegraphics[page=1, trim=0cm 15.8cm 0cm 0cm, clip, width=1.00\textwidth]{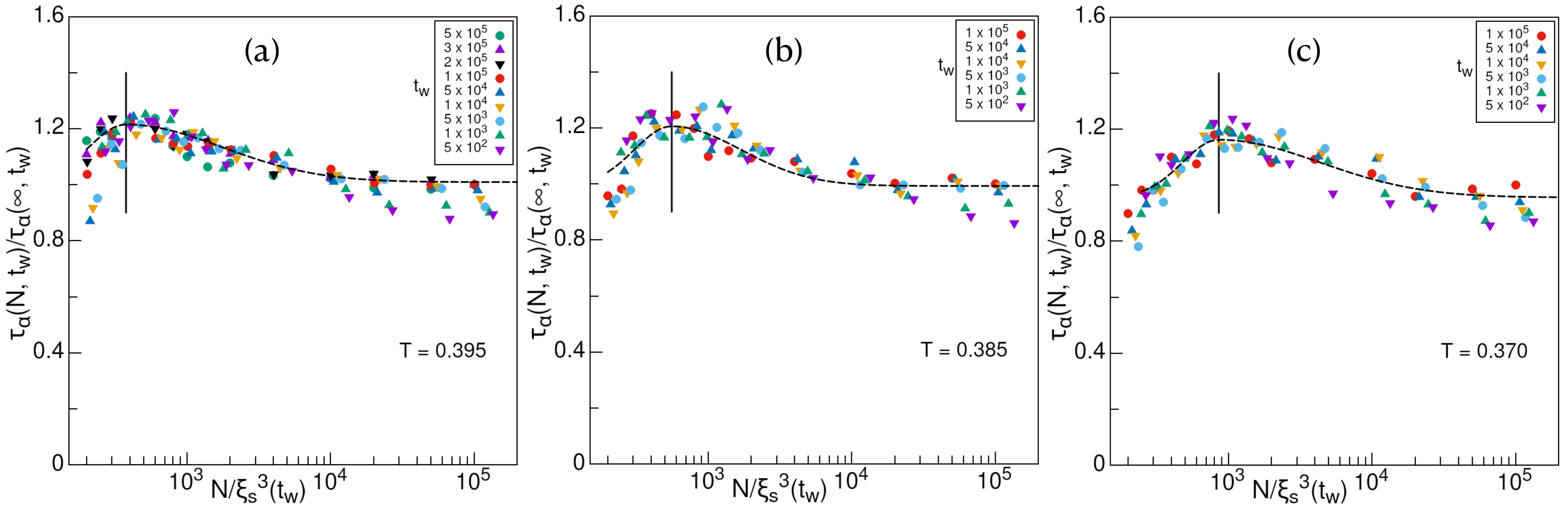}
    \caption{Finite-size scaling of $\alpha$-relaxation time in aging dynamics at temperatures (a) $0.395$, (b) $0.385$, and (c) $0.370$. Collapsed data are fitted with an asymmetric Gaussian function, and the mean of the fitted Gaussian is shown using a vertical straight line. The corresponding $N$ value at the mean is considered as the estimated correlation volume~\cite{smarajit_pre_2012,george_pre_2012} for the maximum $t_w$, and others are rescaled by the scaling factor of $\xi_s(t_w)$. These master curves are presented in a single figure in the main text, by rescaling using the scaled factors of $0.40, 0.63,$ and $1.00$ for $T = 0.395, 0.385,$ and $0.370$ respectively.}
    \label{fss_of_tau}
\end{figure*}

\begin{figure*}[htp!]
    \centering
    \includegraphics[page=1, trim=0cm 5.3cm 0cm 0cm, clip, width=0.67\textwidth]{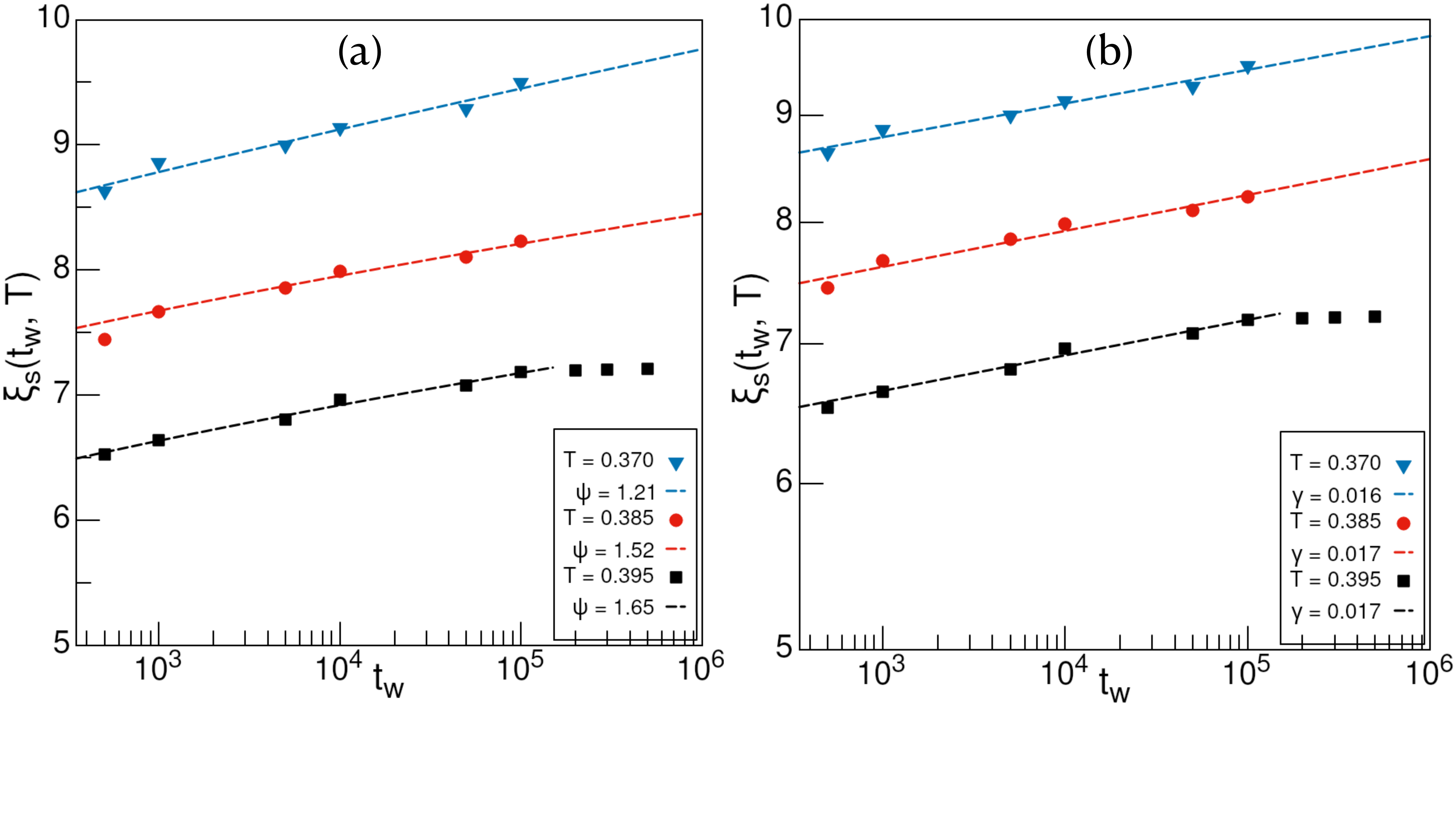}
    \caption{Waiting time dependence of the estimated static length scale $\xi_s(t_w, T)$ for different temperatures is fitted with (a) logarithmic function: $\xi_s(t_w) \sim \xi_{s_0} + [\ln(t_w)]^{1/\psi}$~\cite{rieger_epl_1994,rieger_prb_1996,fisher_prb_1988}, and (b) algebraic function: $\xi_s(t_w) \sim t_w^\gamma$~\cite{rieger_prb_1996}. To best describe the fitting function, figures are represented in (a) linear-log scale and (b) log-log scale. Fitted lines are shown within the aging regime; after that $\xi_s(t_w, T)$ has no waiting time dependency.}
    \label{xis_vs_tw}
\end{figure*}

\clearpage
\section*{B. FSS of Dynamical Heterogeneity}
\begin{figure*}[htp!]
    \centering
    \includegraphics[page=1, trim=0cm 8.8cm 0cm 0cm, clip, width=0.75\textwidth]{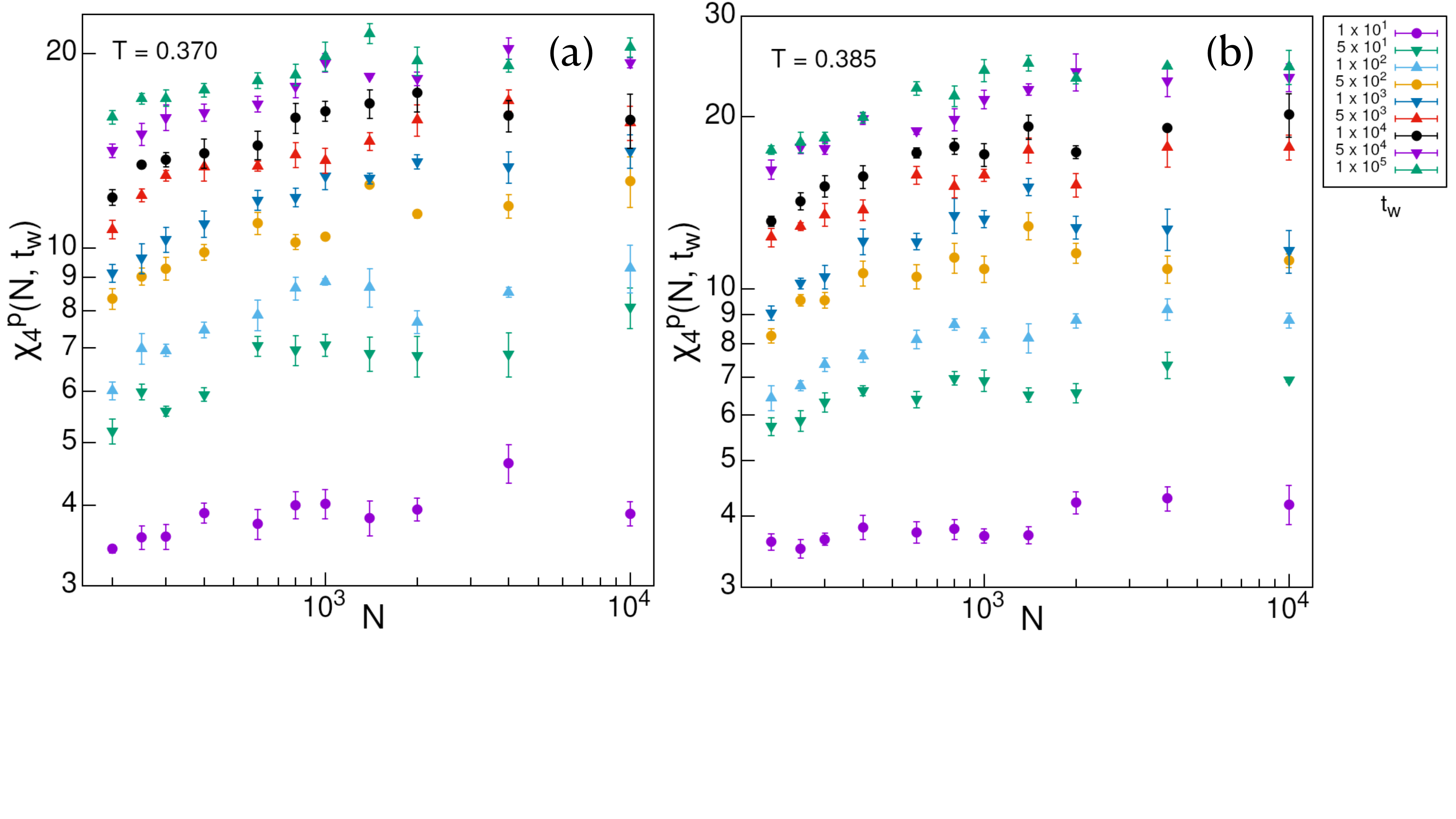}
    \caption{System size dependence of the peak value of dynamical heterogeneity $\chi_4^p(N, t_w)$ for different waiting times is shown for temperatures (a) $0.370$, and (b) $0.385$. It is clearly evident that, for a fixed waiting time, $\chi_4^p(N, t_w)$ is growing with system size, and apparently constant for $N \ge 2000$. Also for a fixed system size, $\chi_4^p(N, t_w)$ is growing with waiting time, and saturates for $t_w \ge \tau_\alpha^e(T)$.}
    \label{chi4-peak_vs_N}
\end{figure*}

\begin{figure*}[htp!]
    \centering
    \includegraphics[page=1, trim=0cm 7.2cm 0cm 0cm, clip, width=0.75\textwidth]{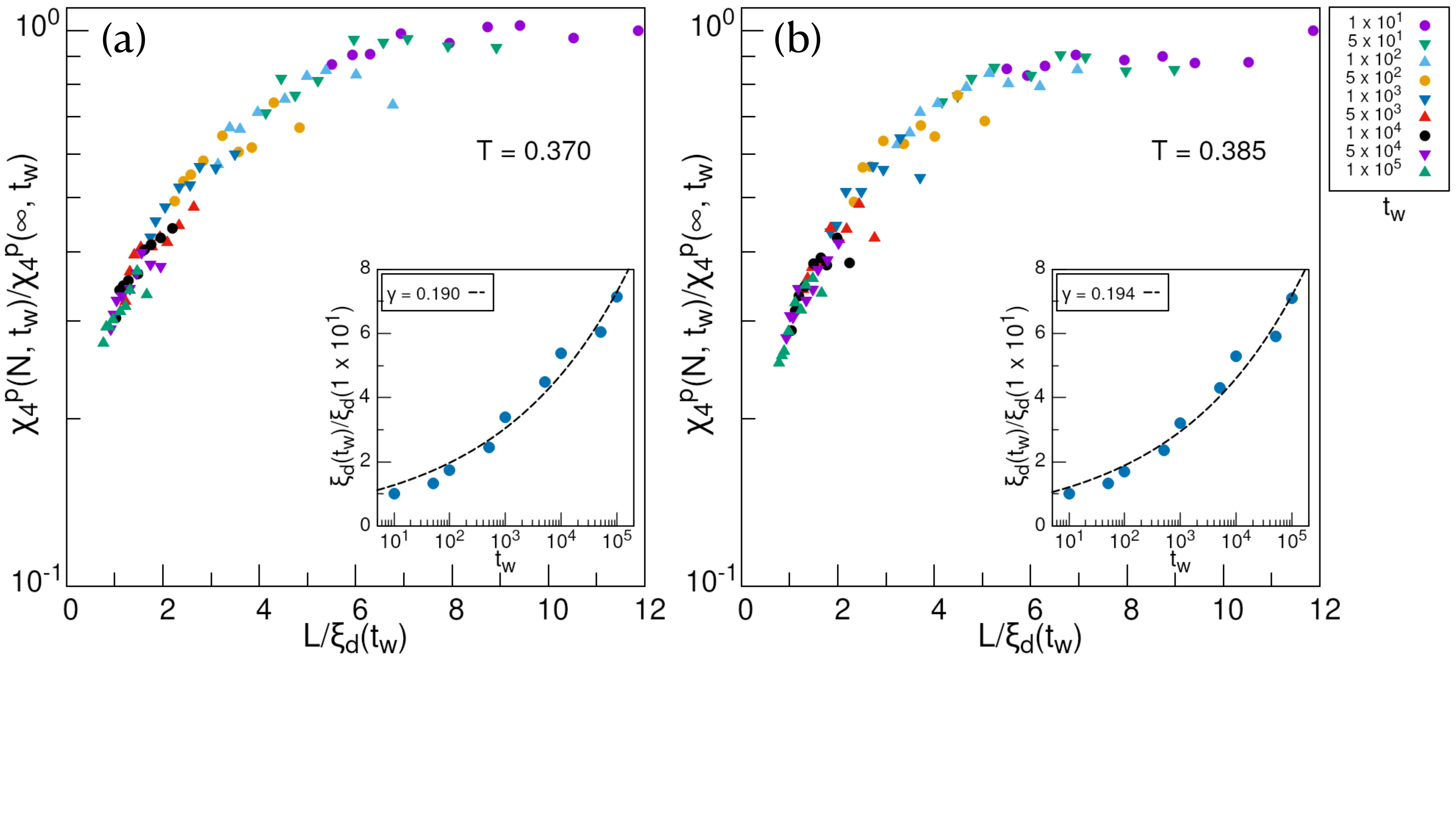}
    \caption{Finite size scaling of $\chi_4^p(N, t_w)$ is performed by rescaling the x-axis and y-axis using a suitable length scale $\xi_d(t_w)$ and $\chi_4^p$ for an infinite system size, respectively, is shown for temperatures (a) $0.370$, and (b) $0.385$. Inset shows growth of the normalized dynamic length scale $\xi_d(t_w)$ with waiting time. These data are best represented by an algebraic fitting function: $\xi_d(t_w) \sim t_w^\gamma$~\cite{parisi_1999,saroj_prl_2012}, and $\gamma$ takes $0.190$ and $0.194$ for temperatures $0.370$ and $0.385$ respectively.}
    \label{fss_of_chi4}
\end{figure*}

\begin{figure*}[htp!]
    \centering
    \includegraphics[page=1, trim=0cm 16.5cm 0cm 0cm, clip, width=1.00\textwidth]{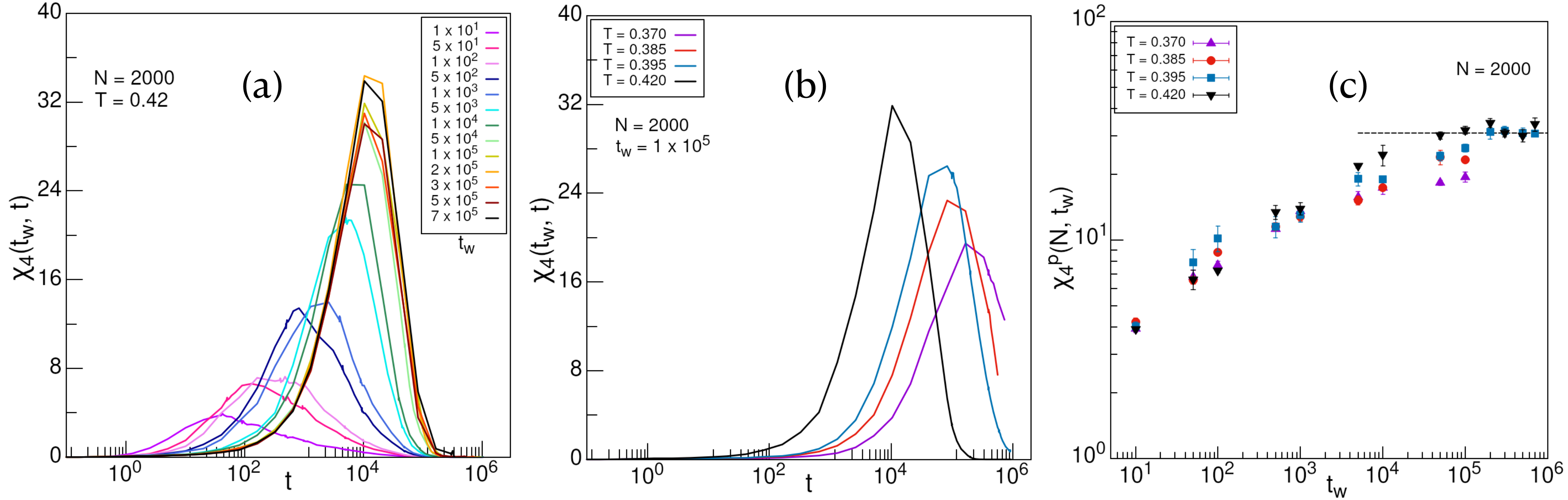}
    \caption{(a) Growth of dynamical heterogeneity $\chi_4(t_w, t)$ with increasing waiting time is shown for $T = 0.420$ and $N = 2000$. For this temperature, $\chi_4^p$ starts to saturate at $t_w \ge 5\times10^4$. It is interesting to observe that after a few smaller values of waiting time ($t_w \ge 1 \times 10^2$), the peak height of $\chi_4$ in the aging regime starts to approach the envelope of the $\chi_4$ in equilibrium ($t_w \ge 5 \times 10^4$). (b) Growth of $\chi_4(t_w, t)$  at a fixed waiting time of $t_w = 1 \times 10^5$ and $N = 2000$ is shown for different temperatures. This clearly shows that, for higher temperatures, $\chi_4(t_w, t)$ has a larger value at a given time as it peaks earlier, consistent with the observation in (a). (c) Waiting time dependence of $\chi_4^p(N, t_w)$ is shown for different temperatures. The value of $\chi_4^p$ seems to saturate at the same height for $T\le T_{MCT}$, after $t_w$ attains $\tau_\alpha^e(T)$.}
    \label{chi4-peak_vs_tw}
\end{figure*}

\begin{figure*}[htp!]
    \centering
    \includegraphics[page=1, trim=0cm 5.8cm 0cm 0cm, clip, width=0.70\textwidth]{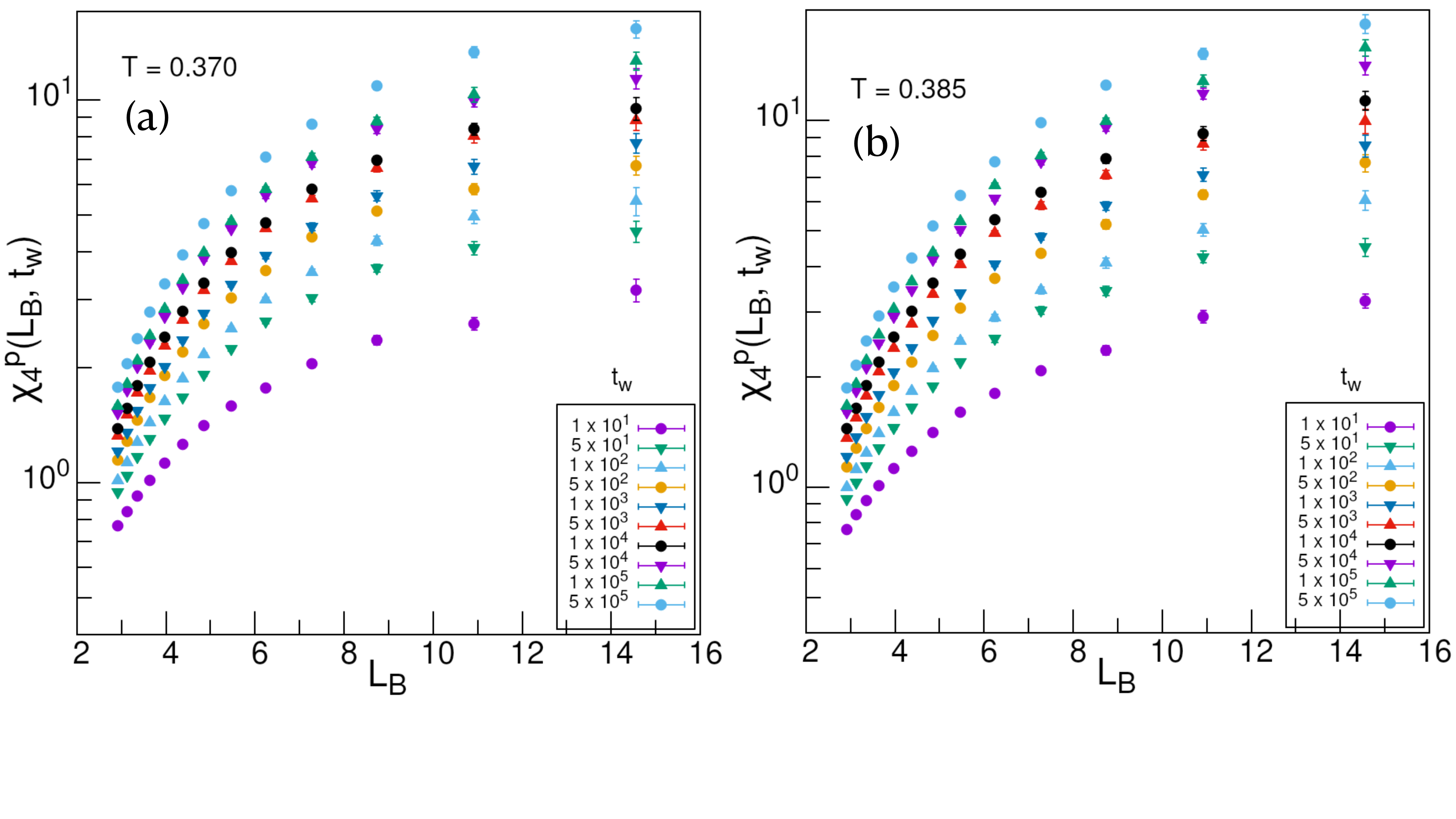}
    \caption{Block size dependence of the peak value of dynamical heterogeneity $\chi_4^p(L_B, t_w)$, calculated on a system size of $N = 100\,000$ for different waiting times is shown for temperatures (a) $0.370$, and (b) $0.385$. It explicitly shows that, for a fixed waiting time, $\chi_4^p(L_B, t_w)$ is growing with block size, and saturates at $\chi_4^p(\infty, t_w)$. It is interesting to note that for a fixed block size, $\chi_4^p(L_B, t_w)$ increases with waiting time, and saturates for $t_w \ge \tau_\alpha^e(T)$.}
    \label{chi4-peak-block_vs_lb}
\end{figure*}

\begin{figure*}[htp!]
    \centering
    \includegraphics[page=1, trim=0cm 8.4cm 0cm 0cm, clip, width=0.78\textwidth]{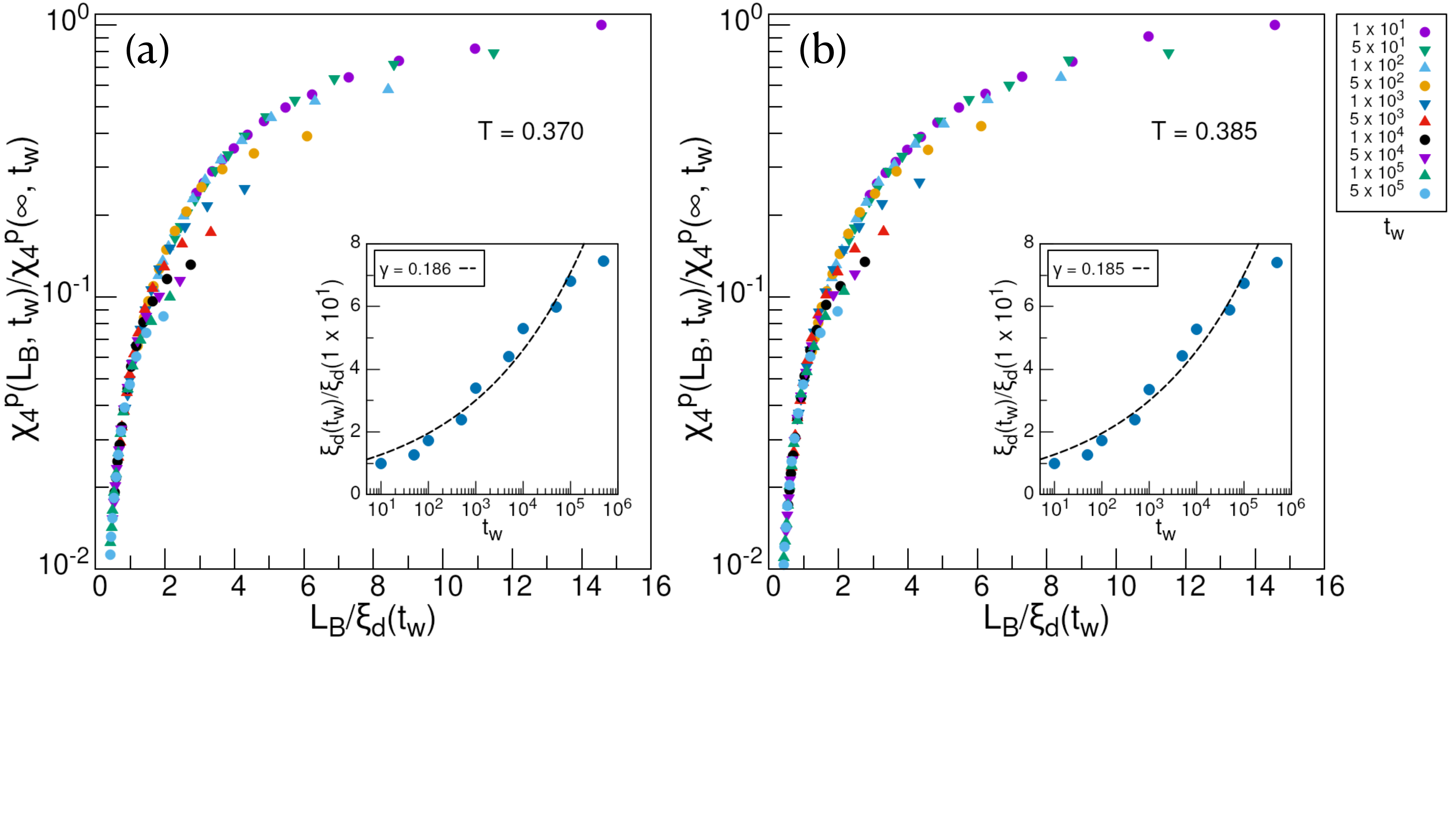}
    \caption{Data collapse is done by rescaling the x-axis using a suitable length scale $\xi_d(t_w)$ and the y-axis using the saturation value of $\chi_4^p$ for an infinite block-size, is shown for temperatures (a) $0.370$, and (b) $0.385$. Inset shows the growth of the normalized dynamic length scale $\xi_d(t_w)$ in variation with waiting time. These data are best described by an algebraic fitting function: $\xi_d(t_w) \sim t_w^\gamma$, and $\gamma$ takes $0.186$ and $0.185$ for temperatures $0.370$ and $0.385$, respectively.}
    \label{fss_of_chi4-block}
\end{figure*}

\begin{figure*}[htp!]
    \centering
    \includegraphics[page=1, trim=0cm 16.5cm 0cm 0cm, clip, width=1.00\textwidth]{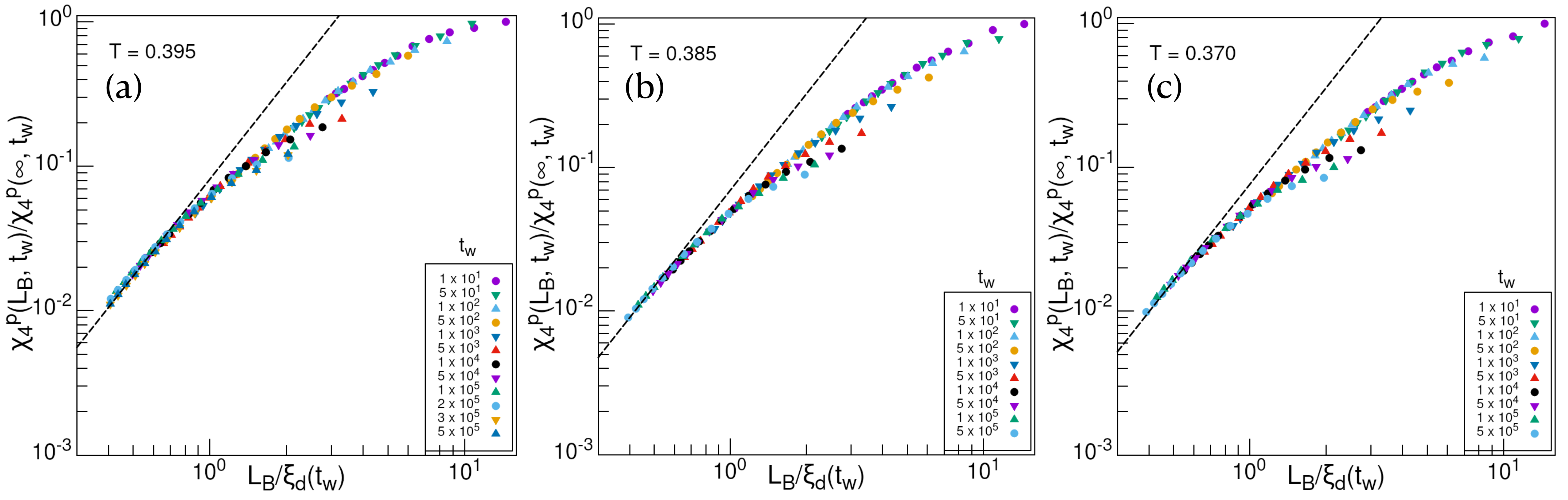}
    \caption{The scaling function $\mathcal{G}(x) \propto x^{2-\eta}$ for $x \rightarrow 0$, is fitted with the collapsed data of $\chi_4^p(L_B, t_w)$ in the block analysis, is shown for temperatures (a) $0.395$, (b) $0.385$, and (c) $0.370$. The exponent $\eta$ takes the value $\textendash 0.23$ for all temperatures, well in agreement with Ref.~\cite{indra_prl_2017}.}
    \label{exponent_fit}
\end{figure*}

\begin{figure*}[htp!]
    \centering
    \includegraphics[page=1, trim=0cm 10.2cm 0cm 0cm, clip, width=0.78\textwidth]{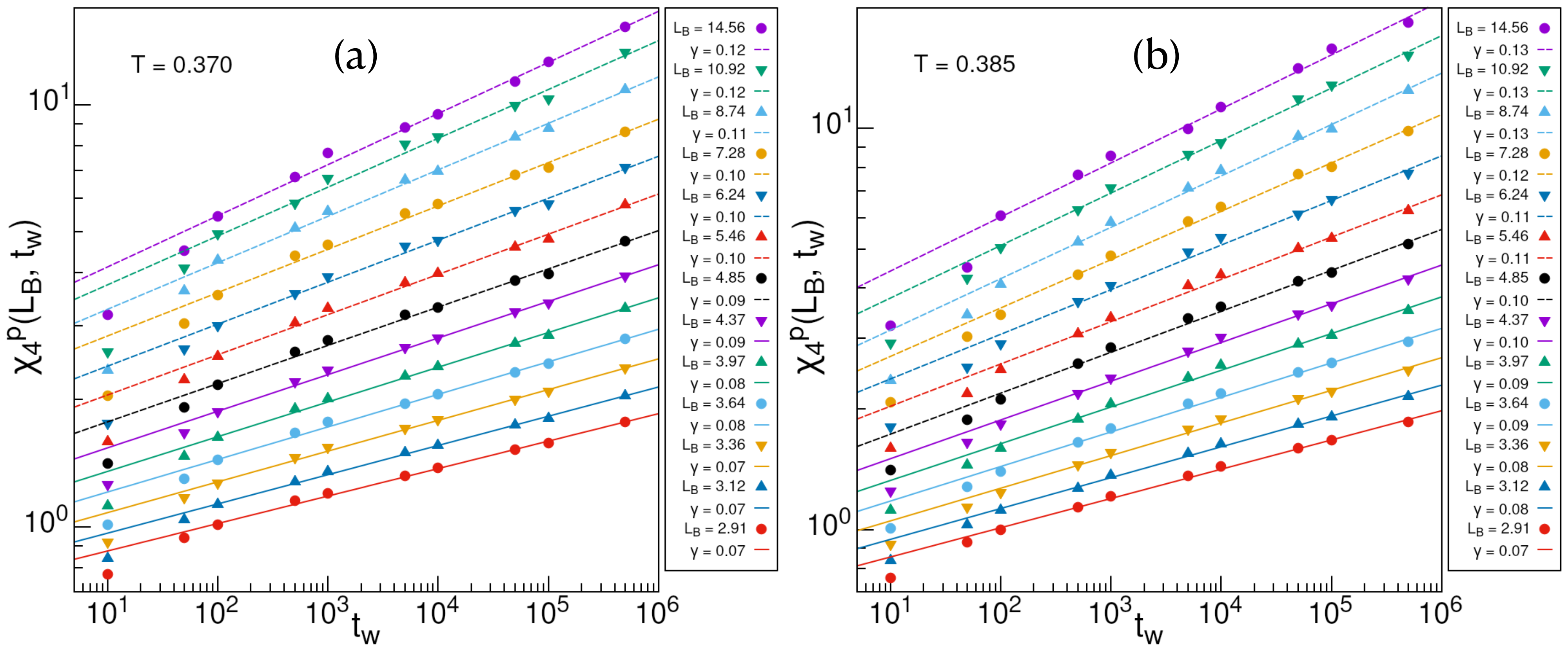}
    \caption{Waiting time dependence of $\chi_4^p(L_B, t_w)$ is shown for different block sizes, and is presented for temperatures (a) $0.370$, and (b) $0.385$. The growth of $\chi_4^p(L_B, t_w)$ is fitted with $t_w^\gamma$ and the exponent $\gamma$ ranges between $0.07\textendash0.12$, and $0.07\textendash0.13$ for temperatures (a) $0.370$, and (b) $0.385$ respectively.}
    \label{chi4-peak-block_vs_tw}
\end{figure*}

\clearpage
\section*{C. Dynamic Length Scale from Spatial Correlation Functions}
\begin{figure*}[htp!]
    \centering
    \includegraphics[page=1, trim=0cm 0cm 5.2cm 0cm, clip, width=1.00\textwidth]{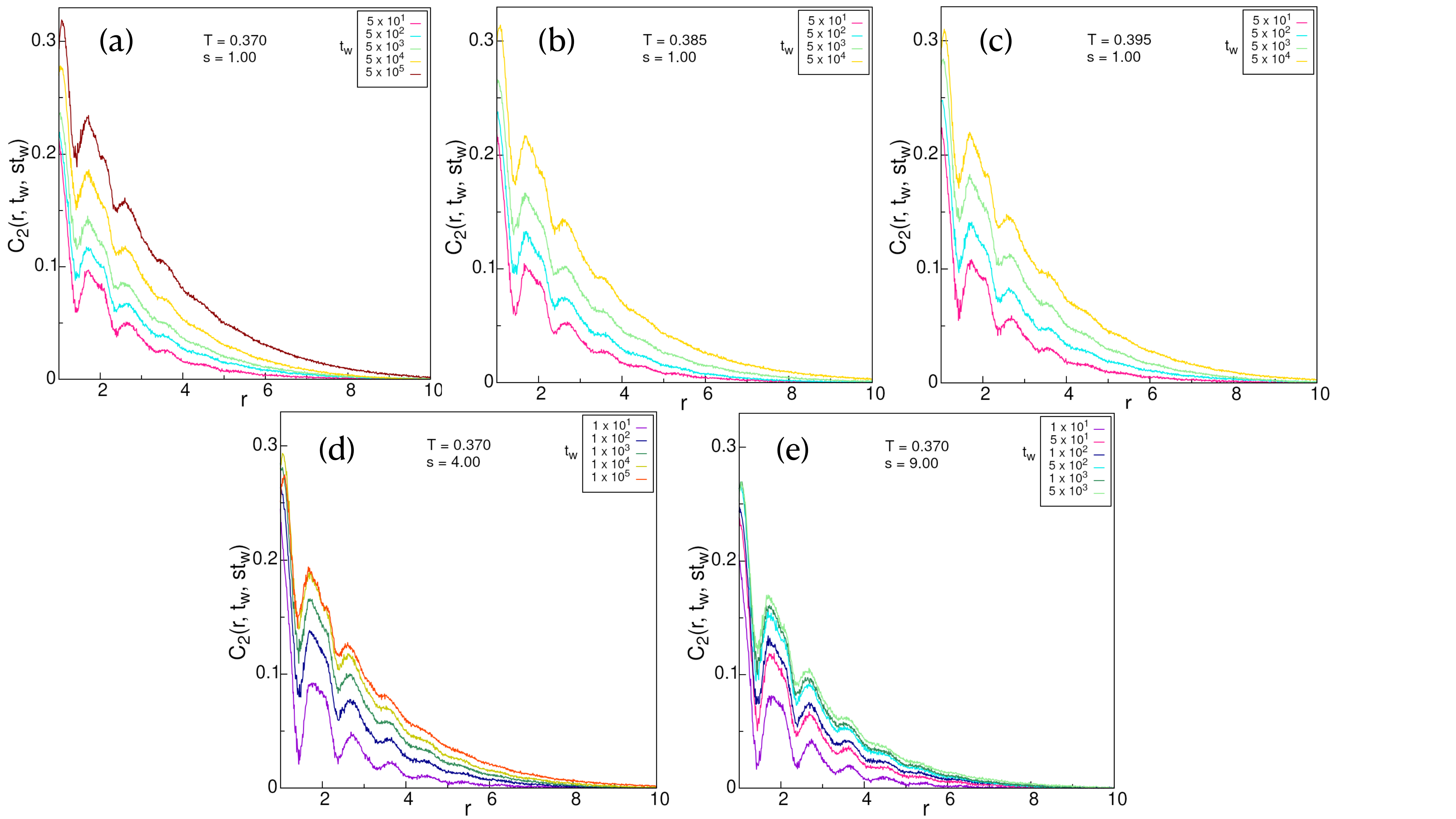}
    \caption{Waiting time dependence of the two-point spatial correlation function is shown for temperatures (a) $0.370$, (b) $0.385$, (c) $0.395$ at $s = 1.00$, and (d) $s= 4.00$, (e) $s = 9.00$ at $T = 0.370$. We computed this correlation function at $\tau_\alpha$ or multiple of $\tau_\alpha$, corresponding to the waiting time by varying the value of $s$. This definition ($t = st_w$)~\cite{parisi_1999} is fairly holds in the aging regime, as $\tau_\alpha$ has a power-law growth of waiting time. It is clearly evident that in (a)$\textendash$(c) for a fixed $t_w$, the integrated area of $C_2(r, t_w, st_w)$ is nearly equal for all studied temperatures, as the simulated temperatures are nearby. However, in (d) and (e) for $s = 4.00$ and $9.00$ at $T = 0.370$, the curves are on top of each other at larger waiting times, where the time difference $t$ is beyond the aging regime.}
    \label{scf-norm}
\end{figure*}

\begin{figure*}[htp!]
    \centering
    \includegraphics[page=1, trim=0cm 0cm 5.2cm 0cm, clip, width=1.00\textwidth]{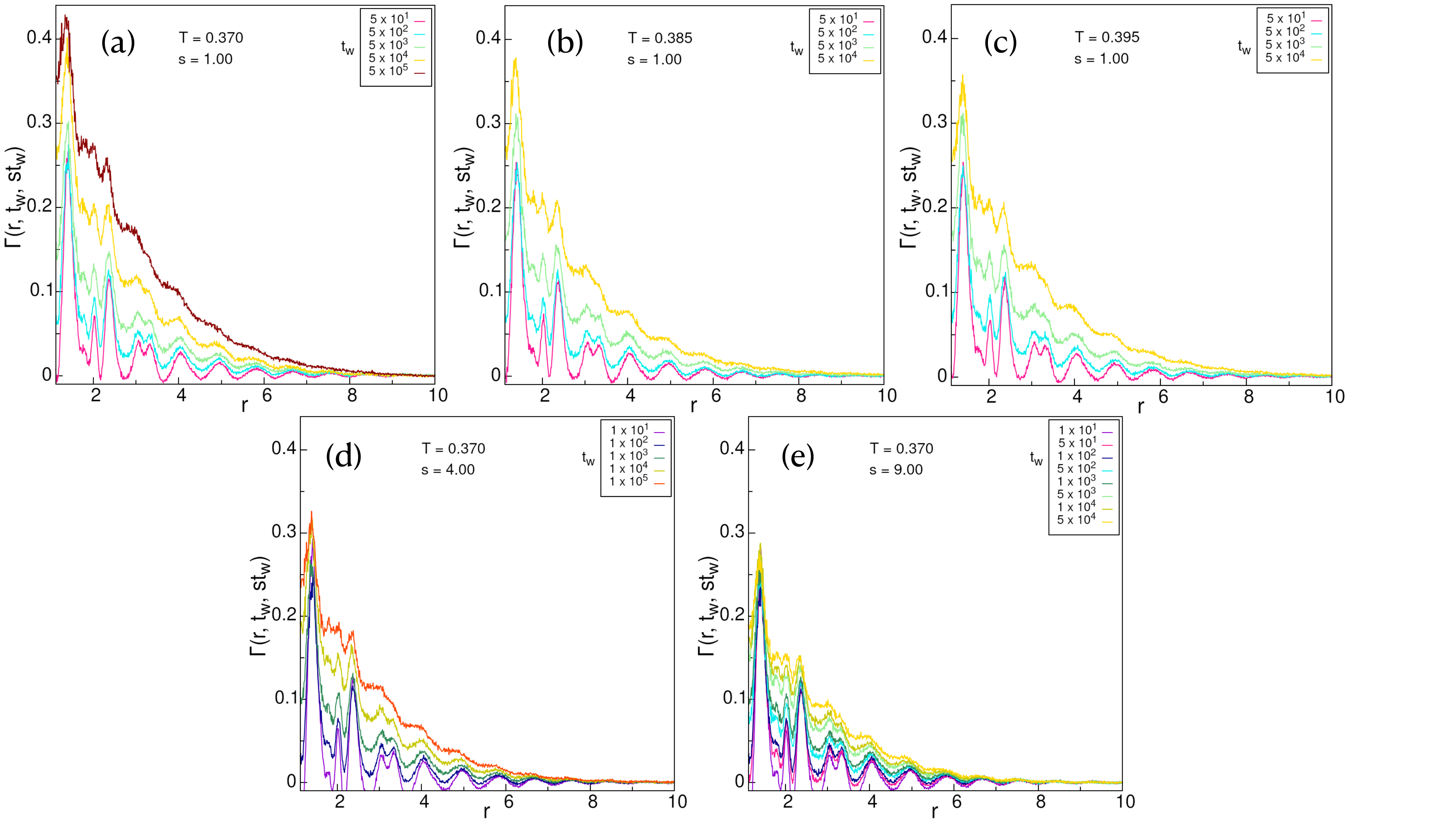}
    \caption{Waiting time dependence of the spatial displacement correlation function is shown for temperatures (a) $0.370$, (b) $0.385$, (c) $0.395$ at $s = 1.00$, and (d) $s= 4.00$, (e) $s = 9.00$ at $T = 0.370$. We also computed this correlation function at $\tau_\alpha$ or multiple of $\tau_\alpha$, corresponding to the waiting time by varying the value of $s$. It is clearly evident that in (a)$\textendash$(c) for a fixed $t_w$, the integrated area of $\Gamma(r, t_w, st_w)$ is nearly equal for all studied temperatures, as the simulated temperatures are nearby. However, in (d) and (e) for $s = 4.00$ and $9.00$ at $T = 0.370$, the curves are on top of each other at larger waiting times, where the time difference $t$ is beyond the aging regime.}
    \label{ddcor-norm}
\end{figure*}

\clearpage
\bibliographystyle{unsrt}
\bibliography{bibliog_supple.bib}